\documentclass[journal,comsoc]{IEEEtran}

\usepackage{graphicx,epsfig}
\usepackage[noadjust]{cite}
\usepackage{mcite}
\usepackage{amsfonts,helvet}
\usepackage{fancyhdr}
\usepackage{threeparttable}
\usepackage{epsf}
\usepackage{amsthm}
\usepackage{amsmath}
\usepackage{siunitx}
\usepackage{amssymb}
\usepackage{booktabs}

\usepackage{dsfont}
\usepackage{subfigure}
\usepackage{color}
\usepackage[linesnumbered,ruled, noend]{algorithm2e}
\usepackage{algpseudocode}
\usepackage{algcompatible}
\usepackage{enumerate}
\usepackage{ragged2e}
\usepackage{algorithm2e}
\usepackage{optidef}

\usepackage{wrapfig}
\usepackage{cancel}

\newtheorem{theorem}{Theorem}

\newtheorem{corollary}{Corollary}

\newtheorem{lemma}{Lemma}

\newtheorem{proposition}{Proposition}
\newtheorem{remark}{Remark}

\newcommand\blfootnote[1]{%
  \begingroup
  \renewcommand\thefootnote{}\footnote{#1}%
  \addtocounter{footnote}{-1}%
  \endgroup
}
\usepackage{eucal}

\usepackage{graphicx,epsfig}
\usepackage[noadjust]{cite}
\usepackage{mcite}
\usepackage{amsfonts,helvet}
\usepackage{fancyhdr}
\usepackage{threeparttable}
\usepackage{epsf}
\usepackage{amsthm}
\usepackage{amsmath}
\usepackage{siunitx}
\usepackage{amssymb}
\usepackage{booktabs}
\usepackage{dsfont}
\usepackage{subfigure}
\usepackage{color}
\usepackage[linesnumbered,ruled, noend]{algorithm2e}
\usepackage{algpseudocode}
\usepackage{algcompatible}
\usepackage{enumerate}
\usepackage{cancel}
\usepackage{multirow} 
\usepackage{threeparttable}

\usepackage{graphicx,epsfig}
\usepackage[noadjust]{cite}
\usepackage{mcite}
\usepackage{amsfonts,helvet}
\usepackage{fancyhdr}
\usepackage{threeparttable}
\usepackage{epsf,epsfig}
\usepackage{amsthm}
\usepackage{amsmath}
\usepackage{siunitx}
\usepackage{amssymb}
\usepackage{booktabs}
\usepackage{dsfont}
\usepackage{subfigure}
\usepackage{color}
\usepackage[linesnumbered,ruled,noend]{algorithm2e}
\usepackage{algpseudocode}
\usepackage{algcompatible}
\usepackage{enumerate}
\usepackage{cancel}
\usepackage{bbm}
\usepackage{graphicx,subfigure}
\usepackage{graphicx}
\usepackage{array}
\newcolumntype{P}[1]{>{\centering\arraybackslash}p{#1}}
\usepackage{eucal}

\def\ba{{\bf a}}

\def\be{{\bf e}}
\def\bff{{\bf f}}
\def\bg{{\bf g}}
\def\bh{{\bf h}}

\def\bn{{\bf n}}

\def\bq{{\bf q}}
\def\br{{\bf r}}
\def\bs{{\bf s}}

\def\bu{{\bf u}}
\def\bv{{\bf v}}
\def\bw{{\bf w}}
\def\bx{{\bf x}}
\def\by{{\bf y}}
\def\bz{{\bf z}}

\def\bA{{\bf A}}
\def\bB{{\bf B}}
\def\bC{{\bf C}}
\def\bD{{\bf D}}

\def\bF{{\bf F}}
\def\bG{{\bf G}}
\def\bH{{\bf H}}
\def\bI{{\bf I}}

\def\bK{{\bf K}}

\def\bM{{\bf M}}
\def\bN{{\bf N}}

\def\bR{{\bf R}}

\def\bV{{\bf V}}
\def\bW{{\bf W}}

\def\cC{\mbox{$\mathcal{C}$}}

\def\cL{\mbox{$\mathcal{L}$}}

\def\cN{\mbox{$\mathcal{N}$}}

\def\bbC{\mbox{$\mathbb{C}$}}

\def\bbE{\mbox{$\mathbb{E}$}}

\begin{document}
\title{Full-Duplex Multiuser MISO Under Coarse Quantization: Per-Antenna SQNR Analysis and Beamforming Design}
% Full-Duplex MU-MISO Systems with Coarse Quantization: How Many Bits Do We Need?
%Full-Duplex MU-MISO Systems with Coarse Quantization: How Many Bits Do We Need?
% Hardware-Efficient Full-Duplex MISO: Optimizing System Performance Under Coarse Quantization and Self-Interference Constraints
\author{Seunghyeong Yoo, Jaehyun Kim, Seokjun Park, Mintaek Oh, Namyoon Lee, and Jinseok Choi
\thanks{S. Yoo, J. Kim, S. Park, M. Oh, and J. Choi are with the Department of Electrical Engineering, Korea Advanced National Institute of Science and Technology (KAIST), Daejeon, 34141, South Korea (e-mail: {\texttt{\{seunghyeong, sury072648,  sj.park, ohmin, jinseok\}@kaist.ac.kr}}).

N. Lee is with the Department of Electrical Engineering, Pohang University of Science and Technology (POSTECH), Pohang 37673, South Korea (e-mail: {\texttt{nylee@postech.ac.kr}}).
}
}

\maketitle
\setcounter{page}{1} 
\begin{abstract}
% 1) FD 시스템이 무선통신에서 각광받고 있음을 작성.
% 2) 이 논문은 어떤 시스템에서 무엇을 제안하는지 작성.
% 3-1) SQNR-based analysis 수행 작성.
% 3-2) analysis를 기반으로 알고리즘 제안 작성. (이전 논문: 무엇이 challenge 인지 작성되었음. 지금은 작성 x)
% 4) 시뮬레이션은 이론적인 analysis와 알고리즘의 우수함을 보인다는 것을 작성.
%% 200 Words %%%
% Full-duplex (FD) communications have emerged as a promising solution to realize the high data rate capability required in wireless networks.
We investigate full-duplex (FD) multi-user multiple-input single-output systems with coarse quantization, aiming to characterize the impact of employing low-resolution analog-to-digital converters (ADCs) on  self-interference (SI) and to develop a quantization- and SI-aware beamforming method that alleviates quantization-induced performance degradation in the FD systems.
We first present an analysis on the per-antenna signal-to-quantization noise ratio for conventional linear beamformers to provide the desired range of the number of analog-to-digital converter (ADC) bits, providing system insights for reliable FD operation in regard to the ADC resolution and beamforming strategy.
Motivated by the insights, we then propose an SI-aware beamforming method that mitigates residual SI and quantization distortion.
The resulting spectral efficiency (SE) maximization problem is decomposed into two tractable subproblems solved via alternating optimization: precoder and combiner design.
The precoder optimization is formulated as a generalized eigenvalue problem, where the dominant eigenvector yields the best stationary solution through power iteration, while the combiner is derived as a quantization-aware minimum mean-squared error (MMSE) filter.
Numerical studies show that the number of required ADC bits with the proposed beamforming falls within the derived theoretical range while achieving the highest SE compared to benchmarks.

\end{abstract}
\vspace{-3mm}

%%%%%%%%%% introduction %%%%%%%%%%
\section{Introduction}
\blfootnote{The conference version of this paper \cite{yoo:vtc:2023} has been published to IEEE Vehicular Technology Conference (VTC).}
% 저해상도 양자화 FD 시스템에 대한 모티베이션
% 1. 현재 및 앞으로의 네트워크에서 요구하는 바에 대해서 작성하여, low-resolution과 FD 시스템이 필요한 이유에 대해 서술. 장점 위주로 작성.
% 2. FD 시스템과 low resolution quantization 이 직면하고 있는 문제들에 대해서 서술. 이후, 이러한 문제를 해결하기 위해 무엇을 이 논문에서 다루려고 하는지 명확하게 작성.
Future wireless networks beyond 5G target lower power consumption, higher data rates, and enhanced reliability \cite{zhang:vtmag:19, Chowdhury:oj-coms:20}, while facing critical challenges in improving spectral (SE) and energy efficiency (EE).
A promising solution combines full-duplex (FD) technology with coarse quantization. 
FD systems enable simultaneous transmission and reception, potentially doubling SE compared to half-duplex (HD) systems \cite{li:commmag:17}.
Additionally, low-resolution quantization can effectively reduce power consumption for multi-antenna systems, thereby increasing EE while minimizing SE degradation \cite{jacobsson:tcom:2017}.

However, FD systems face inherent limitations due to self-interference (SI) at the uplink (UL) receiver and co-channel interference (CCI) affecting downlink (DL) users \cite{li:commmag:17}. 
Quantization errors further complicate the practical implementation of FD systems by introducing coupled distortion--interference effects. 
Therefore, a thorough understanding of FD system design under coarse quantization is crucial to fully realize its potential.
In this context, we present a comprehensive characterization of FD multi-user multiple-input single-output (MU-MISO) system design and propose a beamforming method  under coarse quantization.

% 본 논문의 목적: 1. 우리의 시스템에서 고려해야할 문제가 무엇인지 명확하게 서술 (low-resolution 및 FD 시스템이 직면하는 문제점들) 
% --> 2. low-resolution에 SI가 영향을 주는데 어느정도 영향을 주는지 명확하게 분석하기 위해, SU-MISO 및 MU-MISO에서 linear precoder를 사용하여 low resolution에 주는 SI의 영향을 파악. 
% --> 3. 이렇게 도출한 insights를 기반으로, advanced beamforming을 설계해야함을 어필. 따라서, 어떻게 beamforming을 수행했는지 명확히 서술.
% --> 4. 시뮬레이션 결과에서 이론적으로 derive한 분석과 numerical을 함께 비교하여, 우리의 분석이 타당함을 입증함. 또한, SE(main performance metric) 및 EE 측면에서 우리의 알고리즘의 우수성을 강조함.
%%%%% prior works %%%%% 
% Full-Duplex prior works: SE 측면에 대해서, 도입했던 지난 works들을 작성.
% - FD 시스템에서의 고유한 문제(SI 등)를 해결해보는 시도들을 작성.
\subsection{Related Works}
FD systems have been widely investigated due to their potential to enhance SE compared to HD systems \cite{chen:electlett:1998, riihonen:tsp:2011, duarte:twc:2012, bharadia:sigcomm:13, everett:twc:2014, khaledian:tmtt:2018, kim2024learning, suraweera:twc:2014, almradi:tcom:2018, nguyen:twc:2014, nguyen:twc:2019, shao:access:2019, huberman:twc:2014, kim:tvt:16, cirik:wcl:2015, dai:access:19, kong:twc:2017, anokye2021full, ding:tvt:2020, riihonen2012analog, korpi:twc:2014}.
The main challenge in FD operation, self-interference (SI), has motivated extensive research on SI cancellation (SIC) from hardware and signal processing perspectives \cite{chen:electlett:1998, riihonen:tsp:2011, duarte:twc:2012, bharadia:sigcomm:13, everett:twc:2014, khaledian:tmtt:2018, kim2024learning}.
One of the earliest studies \cite{chen:electlett:1998} demonstrated an experimental radio frequency (RF) system enabling FD operation via SIC.
Subsequent works investigated various SIC strategies, including spatial suppression in FD relay systems \cite{riihonen:tsp:2011}, passive suppression combined with active analog and digital SIC \cite{duarte:twc:2012}, joint analog-digital SIC achieving up to 110 dB of SI attenuation under typical signal-to-noise ratio (SNR) conditions \cite{bharadia:sigcomm:13}. 
In \cite{everett:twc:2014}, experimental evaluations also demonstrated the effectiveness of passive mechanisms such as directional isolation, absorptive shielding, and cross-polarization.
Recent advances include novel analog SIC techniques for single-antenna in-band FD systems \cite{khaledian:tmtt:2018} and machine learning-based digital SIC methods that suppress residual SI down to receiver noise levels, thereby enabling practical FD operation across diverse scenarios \cite{kim2024learning}.

% Beamforming: SE 향상 및 MISO 시너지(SI 및 CCI)
% Residual SI, however, can still persist due to hardware impairments.
% 이제, 위에서 서술한 FD에 내제된 문제를 제거하기 위한 노력들과 함께, 이를 바탕으로 SE를 maximization하기 위한 여러 시도들을 작성.
% single cell, relaying system, RIS, massive 등등 여러 framework에서 FD 시스템을 고려하여 beamforming을 설계한 내용을 작성.
% - 단, low resolution 및 high resolution 상관없이 작성하자.
Multi-antenna communication systems have achieved significant advancements in maximizing SE and EE through sophisticated beamforming techniques \cite{choi:twc:19, choi:tsp:2017, choi:twc:2022, dai:commlett:2021, hu:tcom:2019, kim:iotj:2022, park:twc:23}, which have been widely adapted for FD systems \cite{suraweera:twc:2014, almradi:tcom:2018, nguyen:twc:2014, nguyen:twc:2019, shao:access:2019, huberman:twc:2014, kim:tvt:16, cirik:wcl:2015, dai:access:19, kong:twc:2017, anokye2021full, ding:tvt:2020}.
In \cite{suraweera:twc:2014}, joint precoding and decoding designs with rank-one zero-forcing (ZF) were proposed.
\cite{almradi:tcom:2018} introduced the use of hop-by-hop ZF at both the transmitter and receiver in FD relay communications.
\cite{nguyen:twc:2014} presented two beamforming methods employing a Frank-Wolfe method and sequential parametric convex approximation.
Additionally, \cite{nguyen:twc:2019} proposed a low-complexity joint optimization framework to maximize the sum SE in FD MU-MISO systems. 
The framework jointly optimized half-array antenna mode selection, user assignment, and power control to mitigate SI and CCI.
In \cite{huberman:twc:2014}, a joint precoding and SIC transceiver structure was proposed using sequential convex programming for both single-user and MU-MIMO systems.
In \cite{kim:tvt:16}, the duality between multiple-access and broadcast channels was leveraged to transform the DL into the dual UL for the joint optimization of beamformers.
% In \cite{cirik:wcl:2015}, a joint beamforming method was introduced to enhance SE while considering proportional fairness.

% FD + low ADC 및 DAC --> 기존 work 한계
% 이제는, 기존 FD와 함께, 앞서 다뤘던 HD에서 power를 줄이기 위한 노력을 함께 고려한 FD + low ADC/DAC에 대한 기존 works를 작성.
% 가장 중요한 파트: 이 파트의 기존 works는 무엇을 고려했고, 무엇을 고려하지 않았는지에 대해 명확히 알 필요가 있음. 이를 바탕으로 내 work이 필요한 이유에 대해 명확하게 주장할 수 있음.
% 특히, low resolution을 고려한 FD 시스템에, analog cancellation과 digital cancellation을 각각 나누어 고려한 work이 있을지라도, 우리의 work 처럼.. low-resolution에 영향을 주는 SI에 대해 분석한 후, 시스템에서 필요한 필요한 quantization bits 수를 명확하게 제안한 works가 없다는 것을 어필해야 함.
% 이를 바탕으로, advanced beamforming 알고리즘이 필요하다는 것을 입증하기 위함임.
% 따라서, [low resolution을 고려한 FD 시스템에, analog cancellation과 digital cancellation을 각각 나누어 고려한 work]을 우선적으로 선별하여, 이와 비교하는 것으로 우리의 motivation을 명확하게 설명 필요
To maximize both SE and EE, low-resolution quantizers have also been investigated in FD systems \cite{dai:access:19, kong:twc:2017, anokye2021full, ding:tvt:2020}. 
In \cite{dai:access:19}, maximum ratio transmission (MRT) and maximum ratio combining (MRC) were employed in FD massive MU-MIMO systems with low-resolution analog-to-digital converters (ADCs) and digital-to-analog converters (DACs).
\cite{kong:twc:2017} analyzed the SE for FD MU-MISO relaying systems with low-resolution ADCs using MRT and MRC.
\cite{anokye2021full} evaluated the SE and EE in FD cell-free MU-MISO systems with low-resolution ADCs at both the access point (AP) and DL users.
\cite{ding:tvt:2020} analyzed the SE and EE of FD massive MU-MISO systems with low-resolution quantization under a Rician fading channel.
These studies indicated that coarse quantization can enhance EE in FD systems while only marginally degrading SE.

Prior works \cite{dai:access:19, kong:twc:2017, anokye2021full, ding:tvt:2020}  demonstrated the benefits of combining FD techniques and coarse quantization with conventional linear beamformers.
While these studies highlighted potential gains, their reliance on linear beamforming techniques remains suboptimal.
Particularly in the medium-to-high SNR regime, where quantization errors become more pronounced \cite{kong:twc:2017, ding:tvt:2020}, low-resolution quantization poses challenges for traditional linear beamforming in effectively managing quantization errors.
Moreover, these works \cite{dai:access:19, kong:twc:2017, anokye2021full, ding:tvt:2020} did not distinguish the impacts of analog and digital SIC on ADC quantization, while numerical investigations without beamforming revealed that the performance of the FD system with coarse quantization is more limited by analog and digital SIC than the ADC dynamic range \cite{riihonen2012analog, korpi:twc:2014}.
As discussed, comprehensive characterization of FD beamforming systems with coarse quantization, including the impact of the SI to the ADC resolution and advanced FD beamforming strategy, has not been fully explored.

%%%%% 내 연구의 목적 및 논문의 구성 내용 설명 %%%%%
\subsection{Contributions}
In this paper, we investigate FD MU-MISO systems with low-resolution quantizers.
Key contributions are:

% 논문의 contribution:
% 1. Problem Formulation: System modeling
% - low-resolution을 고려한 FD 시스템에, 명확하게 analog SIC와 digital SIC를 나누어 고려했으며, 특히, 우리의 시스템이 어떠한 점에서 풀기 어려운지 어필해야함.
% 2. SQNR-based Analysis: 
% - 결론적으로 ADC bits에 대한 SI 영향이 가장 큰 문제이기 때문에... 이 영향을 우선적으로 linear precoder로 분석하여, 타겟 SQNR에 따른 ADC bits에 대한 이론적인 bound 를 제시함. 이를 통해, 정확히 어느정도의 ADC bit가 필요한지 closed form으로 도출한다는 것을 어필.
% 3. Algorithm (MIMO): formulation challenge solution.
% - 위에서의 인사이트를 바탕으로 advanced BF가 필요함을 강조. BF 방법은 GPI를 이용하는 것이며, 이에 대한 내용을 명확하게 서술. (문제를 generalized eigenvalue problem으로 볼 수 있고... 이에 대한 KKT stationarity condition을 만족하는 해를 구한다는 것...)
% 4. Simulations (Numerical findings): 각 figure 당 하나씩 작성.
% - 특히, 3가지를 어필해야함.
% - 4.1) 이론적으로 보였던 low resolution에 대한 SI의 영향을 SQNR과 bits에 대해 보인 것을 numerical에서도 보였는데, 동일함을 보이며 이론의 타당성을 강조.
% - 4.2) SE 측면에서 제안된 알고리즘의 강점을 서술.
% - 4.3) EE 측면에서도 제안된 알고리즘이 상당한 성능을 보임. 특히 SE와 EE모든 면에 대해서 우수한 성능을 보이기 위해서는 대략적으로 어느 정도의 비트가 필요하다는 것을 어필함.
\begin{itemize}
    % System modeling (다른 논문과는 달리 analog SIC를 분리하여 고려)
    \item In contrast to prior works \cite{dai:access:19, kong:twc:2017, anokye2021full, ding:tvt:2020}, our system model explicitly incorporates the effect of analog SIC on ADC quantization errors.
    Since residual analog SI contributes to ADC quantization errors which directly affects the quality of the UL signals, the influence of analog SIC needs to be considered separately from digital SIC to accurately characterize quantized FD systems.
 
    \item
    % SQNR-based anlysis에 대한 설명
    We then provide analytical insights into the interplay between the residual SI after analog SIC and quantization errors with conventional linear precoders.
    Through the derivation of the average per-antenna signal-to-quantization-noise ratio (SQNR), we show that the desired number of ADC bits logarithmically increase with the residual analog SI to attain a target SQNR.
    This result theoretically quantifies the impact of the residual analog SI on reducing the ADC resolution in the FD system, which is not the case in the HD system.
    We further identify the range of the required number of ADC bits with linear precoders.

    % Algorithm (MISO formulation challenge solution)
    \item Noting that the required number of ADC bits for an optimal precoder which jointly balances SI suppression and DL signal enhancement for sum SE maximization should lie within the derived range, we propose an advanced SI-aware beamforming design that jointly accounts for residual SI, SIC, quantization errors, inter-user interference (IUI), and CCI.
    To this end, we formulate a sum SE maximization problem and decompose it into two subproblems: precoder design and combiner design.
    The precoder optimization is formulated as a generalized eigenvalue problem, where the dominant eigenvector yields the best stationary solution through power iteration, while the combiner is derived as a quantization-aware minimum mean-squared error (MMSE) filter.

    % Numerical results에 대한 설명
    \item In simulations we compare the proposed algorithm with benchmarks in terms of UL average SQNR and evaluate their SE and EE performance.
    Simulation results show that the proposed algorithm effectively mitigates residual SI and quantization distortion under coarse quantization.
    As the ADC resolution increases and quantization errors become negligible, the objective of beamforming gradually shifts toward maximizing the DL SE, deviating from SI suppression, which aligns with the derived insights in our analysis.
    Furthermore, the result reveals that employing 6-7 ADC bits is desirable for the considered FD system in terms of the EE, which requires more bits than the commonly known 4-5 bits in the HD systems.
\end{itemize}

%%%%% Notation %%%%%
$\mathit{Notations}$: $a$ is a scalar, $\ba$ is a vector, $\bA$ is a matrix, $\bI_{N}$ is an identity matrix of size $N \times N$, $\mathbf{0}_{M\times N}$ is a zero matrix of size $M \times N$, $\bA \otimes \bB$ is the Kronecker product of two matrices $\bA$ and $\bB$, $\be_{k}$ is the $k$th standard basis column vector, whose $k$th entry is one and all other entries are zero, and $\cC\cN(\mu,\sigma^{2})$ is a complex Gaussian distribution with mean $\mu$ and variance $\sigma^{2}$.
The operations $(\cdot)^{\sf T}$, $(\cdot)^{\sf H}$, $(\cdot)^{-1}$, $\bbE[\cdot]$, $\mathrm{Tr}(\cdot)$, and $\CMcal{O}(\cdot)$ denote a matrix transpose, Hermitian, matrix inverse, expectation, trace, and big-O operator, respectively.
We use $\mathrm{diag}\{{\bf a}\}$ for a diagonal matrix with elements of $\bf a$, $\mathrm{vec}(\bA)$ for a vectorization of the matrix $\bA$, and $\bA_{(i, j)}$ for its $(i, j)$th component.

%%%%% system model %%%%%
\section{System model}
\label{sec:system}
\begin{figure}[t]
    \centering
    \includegraphics[width=0.9\linewidth]{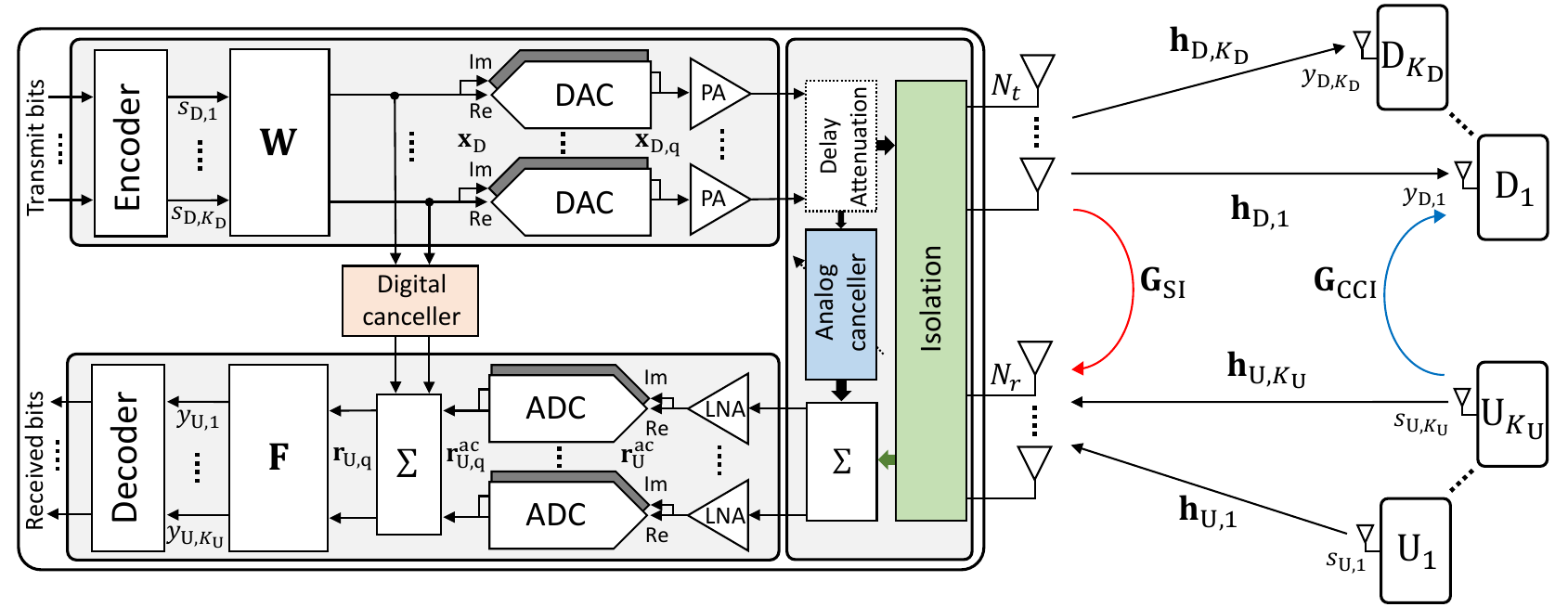}
    \caption{\label{fig:system_model}Full-duplex MU-MISO system model with coarse quantization.}
\end{figure}

We investigate a single-cell FD MU-MISO system with low-resolution quantization. 
In this system, a FD AP equipped with $N_t$ transmit and $N_r$ receive antennas simultaneously serves $K_{\sf D}$ DL and $K_{\sf U}$ UL single-antenna users, as illustrated in Fig.~\ref{fig:system_model}.
Both the ADCs and DACs at the AP operate with a limited number of quantization bits, resulting in non-negligible quantization errors which affect the FD signal processing.

%---- DL 및 UL network 모델링 ----%
On the DL, the AP transmits an independent data symbol $s_{{\sf D},k} \! \sim \! \cC\cN(0,P_{\sf D})$ to the $k$th DL user for $k \! \in \! \{1,\dots,K_{\rm D}\}$, where $P_{\sf D}$ is the AP's maximum transmit power.
Defining $\bs_{\sf D} \! = \! [s_{{\sf D},1}, ..., s_{{\sf D},K_{\sf D}}]^{\sf T} \! \in \! \bbC^{K_{\sf D}}$, a transmit signal vector is given by
\begin{equation}
    \bx_{\sf D} = \bW \bs_{\sf D},
    \label{eq:x_D}
\end{equation}
where $\bW \in \bbC^{N_t \times K_{\sf D}}$ is a precoder, and its $k$th column $\bw_k \in \bbC^{N_t}$ corresponds to the $k$th DL user.
Subsequently, $\bx_{\sf D}$ is processed by low-resolution DACs.
Let $b_{{\sf DAC},m}$ denote the quantization bits of the $m$th DAC.
For analytical tractability, we adopt the additive quantization noise model (AQNM) \cite{fletcher:jstsp:07}, which is a linear approximation of the quantizer \cite{dai:access:19, kong:twc:2017, choi:tsp:2017, choi:twc:2022, dai:commlett:2021, hu:tcom:2019, kim:iotj:2022, park:twc:23}.
Accordingly, the quantized transmit signal is
\begin{align}
    Q(\bx_{\sf D}) \approx \bx_{\sf D,q} = \; \boldsymbol{\Phi}_{\alpha_{\sf DAC}} \bx_{\sf D} + \bq_{\sf DAC},
    \label{eq:x_Dq}
\end{align}
where $Q(\cdot)$ is a scalar quantizer that operates independently on the real and imaginary components, $\boldsymbol{\Phi}_{\alpha_{\sf DAC}} \! = \! \mathrm{diag}\{\alpha_{\sf{DAC},1}, ..., \alpha_{{\sf DAC},N_t}\} \! \in \! \bbC^{N_t \times N_t}$ is a diagonal matrix of DAC quantization loss for the $m$th DAC pair $\alpha_{{\sf DAC},m}$, and
$\bq_{\sf {DAC}} \! \in \! \bbC^{N_t}$ is a DAC quantization noise vector.
Here, $\alpha_{{\sf DAC},m}$ is defined as $\alpha_{{\sf DAC},m} \! = \! 1 - \beta_{{\sf DAC},m} \! \!\in \!\! (0,1)$, where $\beta_{{\sf DAC},m}$ is the normalized mean-squared quantization error, i.e., $\beta_{{\sf DAC},m} \! = \! \frac{\bbE[|x-Q_n(x)|]^2}{\bbE[|x|]^2}$ \cite{fletcher:jstsp:07}.

For $ b_{\sf DAC} \! \leq \! 5$,  $\alpha_{{\sf DAC},m}$ can be found from Table 1 in~\cite{fan:commlett:2015}; for $ b_{\sf DAC} \! > \! 5$, $\beta_{{\sf DAC},m}$ is approximated by $\beta_{{\sf DAC},m} \! \approx \! \frac{\pi \sqrt3}{2}  2^{-2 b_{\sf DAC}}$ \cite{fan:commlett:2015}.
We assume $\bq_{\sf {DAC}} \! \sim \! \cC\cN (\mathbf{0}_{N_t \times 1}, \bR_{\bq_{\sf {DAC}}\bq_{\sf {DAC}}}(\bW))$ whose covariance matrix is~\cite{fletcher:jstsp:07,park:twc:23}:
\begin{align}
   \bR_{\bq_{\sf {DAC}}\bq_{\sf {DAC}}}({\bf W}) =&\; \boldsymbol{\Phi}_{\alpha_{\sf DAC}} \boldsymbol{\Phi}_{\beta_{\sf DAC}} \mathrm{diag}{\left\{ \bbE \left[\bx_{\sf D}\bx_{\sf D}^{\sf H}\right] \right\}} \nonumber \\
    =& \; \boldsymbol{\Phi}_{\alpha_{\sf DAC}} \boldsymbol{\Phi}_{\beta_{\sf DAC}} \mathrm{diag}{\left\{ P_{\sf D} \bW \bW^{\sf H} \right\}}. \label{eq:R_qDACqDAC}
\end{align}

% ---------- figure: system model -----------
% -------------------------------------------

On the UL, the $k$th transmits a symbol $s_{{\sf U},k} \! \sim \! \cC\cN (0, P_{\sf U})$ for $k \! \in \!\! \{1,...,K_{\sf U} \}$, where $P_{\sf U}$ is the maximum transmit power of the UL user.
Let $\bs_{\sf U} \! = \! [s_{{\sf U},1}, ..., s_{{\sf U},K_{\sf U}}]^{\sf T} \! \in \! \bbC^{K_{\sf U}}$ be the vector of $K_{\sf U}$ UL user symbols.
According to the FD systems, the UL analog received signal after analog SIC is expressed as~\cite{kim2024learning}
\begin{align}
    \br_{\sf U}^{\sf ac} = \bH_{\sf U}\bs_{\sf U} + \bG_{\sf SI}^{\sf H}\bx_{\sf D,q} + \bn_{\sf U},
    \label{eq:r_U}
\end{align}
where $\bH_{\sf U} \! = \! \left[{\bh_{{\sf U},1}}, ..., {\bh_{{\sf U},K_{\sf U}}}\right] \! \in \! \bbC^{N_r \times K_{\sf U}}$ is a UL channel matrix, $\bG_{\sf SI} \! = \! \left[\bg_{{\sf SI},1}, ..., \bg_{{\sf SI}, N_r}\right] \! \in \! \bbC^{N_t \times N_r}$ is a SI channel matrix, and $\bn_{\sf U} \! \sim \! \cC\cN(\mathbf{0}_{N_r \times 1},\sigma_{\sf U}^2 \bI_{N_r})$ is a UL additive noise vector.
We assume that each entry of $\bG_{\sf SI}$ follows an independent and identically distributed (IID) $ \cC\cN({0}, \kappa_{\sf a})$, based on the assumption that any line-of-sight component is effectively reduced \cite{ngo2014multipair}.
Here, $\kappa_{\sf a}$ quantifies the analog SIC capability, encompassing passive and active suppression.
The UL channel vector for the $k$th user is modeled as $\bh_{{\sf U},k} \! = \! \sqrt{\rho_{{\sf U},k}} \tilde{\bh}_{{\sf U},k}$,
where $\rho_{{\sf U},k}$ is the UL large-scale path-loss component, and $\tilde{\bh}_{{\sf U},k} \sim \cC\cN(\mathbf{0}_{N_r \times 1},\bC_{{\sf U},k})$ is a UL small-scale fading vector.

After analog SIC, the RF receive vector in \eqref{eq:r_U} is processed by low-resolution ADCs at the AP. 
We denote the number of quantization bits used by the $n$th ADC pair as $b_{{\sf ADC},n}$.
Using the AQNM, the quantization output of the ADCs is given as
\begin{align}
    \nonumber
    Q(\br_{\sf U}^{\sf ac})& \approx \br_{\sf U,q}^{\sf ac} = \; \boldsymbol{\Phi}_{\alpha_{\sf ADC}}\br_{\sf U}^{\sf ac} + \bq_{\sf ADC} 
    \\
    &= \boldsymbol{\Phi}_{\alpha_{\sf ADC}}\bH_{\sf U}\bs_{\sf U} \!+\! \boldsymbol{\Phi}_{\alpha_{\sf ADC}}\bG_{\sf SI}^{\sf H}\bx_{\sf D,q} \!+\! \boldsymbol{\Phi}_{\alpha_{\sf ADC}}\bn_{\sf U} \!+\! \bq_{\sf ADC}, \label{eq:r_ac_Uq}
\end{align}
where the AQNM model parameters for ADC quantization such as $\boldsymbol{\Phi}_{\alpha_{\sf ADC}} \!\! = \! \mathrm{diag}\{\alpha_{\sf{ADC},1}, ..., \alpha_{{\sf ADC},N_r} \! \} \!\! \in \! \bbC^{N_r \times N_r}$, $\beta_{\sf ADC}$, and ${\bf q}_{\sf ADC}$ are defined similarly as the DAC case.
We also assume that $\bq_{\sf {ADC}} \! \sim \! \cC\cN(\mathbf{0}_{N_r \times 1}, \bR_{\bq_{\sf {ADC}}\bq_{\sf {ADC}}}({\bf W}))$ whose covariance matrix is given by~\cite{fletcher:jstsp:07,park:twc:23}:
\begin{align}
    \bR_{\bq_{\sf {ADC}}\bq_{\sf {ADC}}}({\bf W}) = \boldsymbol{\Phi}_{\alpha_{\sf ADC}} \boldsymbol{\Phi}_{\beta_{\sf ADC}} \mathrm{diag}{\left\{ \bbE \left[\br_{\sf U}^{\sf ac}(\br_{\sf U}^{\sf ac})^{\sf H} \right] \right\}},
    \label{eq:R_qADCqADC}
\end{align}
where  
\begin{align}
    \bbE \left[\br_{\sf U}^{\sf ac} (\br_{\sf U}^{\sf ac})^{\sf H} \right] 
    =& \; P_{\sf U}\bH_{\sf U}\bH_{\sf U}^{\sf H} + P_{\sf D} \bG_{\sf SI}^{\sf H} \boldsymbol{\Phi}_{\alpha_{\sf DAC}}\bW\bW^{\sf H} \boldsymbol{\Phi}_{\alpha_{\sf DAC}}^{\sf H} \bG_{\sf SI} \nonumber\\
    &+ \bG_{\sf SI}^{\sf H} \bR_{\bq_{\sf DAC}\bq_{\sf DAC}}({\bf W}) \bG_{\sf SI} + \sigma_{\sf U}^{2}\bI_{N_r}.
    \label{eq:expected_r_U}
\end{align}

After digital SIC \cite{kim2024learning}, the quantized signal in \eqref{eq:r_ac_Uq} becomes
\begin{align}
     \label{eq:r_Uq}
    {\br}_{{\sf U,q}} \!=\! \boldsymbol{\Phi}_{\alpha_{\sf ADC}}\bH_{\sf U}\bs_{\sf U} \!+\! \sqrt{\kappa_{\sf d}}\boldsymbol{\Phi}_{\alpha_{\sf ADC}}\!\bG_{\sf SI}^{\sf H}\bx_{\sf D,q} \!+\! \boldsymbol{\Phi}_{\alpha_{\sf ADC}}\bn_{\sf U} \!+\! \bq_{\sf ADC},
\end{align}
where $\kappa_{\sf d}$ denotes the digital SIC capability.
We note that the digital SIC cannot reduce the quantization noise ${\bf q}_{\sf ADC}$ whose power is proportional to the residual analog SI and inversely proportional to the quantization resolution.
Subsequently, the vector ${\br}_{{\sf U,q}}$ is combined via  $\bF  = [\bff_1,\dots, \bff_{K_{\sf U}} ]\in \bbC^{N_r \times K_{\sf U}} $ as
\begin{align}
    \by_{\sf U}
    =& \; \bF^{\sf H} \boldsymbol{\Phi}_{\alpha_{\sf ADC}} \bH_{\sf U} \bs_{\sf U} +  \sqrt{\kappa_{\sf d}} \bF^{\sf H} \boldsymbol{\Phi}_{\alpha_{\sf ADC}} \bG_{\sf SI}^{\sf H} \boldsymbol{\Phi}_{\alpha_{\sf DAC}} \bW \bs_{\sf D} \nonumber\\
    &+\! \sqrt{\kappa_{\sf d}} \bF^{\sf H} \boldsymbol{\Phi}_{\alpha_{\sf ADC}} \bG_{\sf SI}^{\sf H} \bq_{\sf DAC} \!+\! \bF^{\sf H} \boldsymbol{\Phi}_{\alpha_{\sf ADC}} \bn_{\sf U}\! +\! \bF^{\sf H} \bq_{\sf ADC}.
    \label{eq:y_U}
\end{align}

\begin{remark}
    \label{rm:remark1}
    \normalfont Unlike the HD system, the ADC quantization error in the FD system is dominated by the DL transmit power, analog SIC capability, and precoder, since the residual analog SI power at the receiver is much higher than both the received UL desired signal and UL noise power.
    Thus, it is desirable to carefully determine the number of quantization bits and design the precoder to 1) reduce the residual analog signal power level to lie within the ADC dynamic range, 2) maintain the ADC quantization noise power below the UL desired signal power, and 3) alleviate the coarse quantization-induced distortion.
\end{remark}

Returning to the DL perspective, we represent the received signal vector at the users as
\begin{align}
    \by_{\sf D} 
    =& \; \bH^{\sf H}_{\sf D} \bx_{\sf D,q} + \bG_{\sf CCI}^{\sf H} \bs_{\sf U} + \bn_{\sf D} 
    \nonumber\\
    =& \; \bH^{\sf H}_{\sf D} \boldsymbol{\Phi}_{\alpha_{\sf DAC}} \bW \bs_{\sf D} + \bH^{\sf H}_{\sf D} \bq_{\sf DAC} + \bG_{\sf CCI}^{\sf H} \bs_{\sf U} + \bn_{\sf D},
    \label{eq:y_D}
\end{align}
where $\bH_{\sf D} \! = \! \left[{\bh_{{\sf D},1}}, ..., {\bh_{{\sf D},K_{\sf D}}}\right] \! \in \! \bbC^{N_t \times K_{\sf D}}$ is a DL channel matrix, $\bG_{\sf CCI} \! = \! \left[\bg_{{\sf CCI},1}, ..., \bg_{{\sf CCI},K_{\sf D}}\right] \! \in \! \bbC^{K_{\sf U} \times K_{\sf D}}$ is a CCI channel matrix, and $\bn_{\sf D} \! \sim \! \cC\cN(\mathbf{0}_{K_{\sf D} \times 1},\sigma_{\sf D}^2 \bI_{K_{\sf D}})$ is a DL additive noise vector.
Similar to the UL channel $\bh_{{\sf U},k}$, the DL channel vector for the $k$th user is modeled as $\bh_{{\sf D},k} \! = \! \sqrt{\rho_{{\sf D},k}} \tilde{\bh}_{{\sf D},k}$,
where $\rho_{{\sf D},k}$ is the DL large-scale path-loss component, and $\tilde{\bh}_{{\sf D},k} \! \sim \! \cC\cN({\bf 0}_{N_t \times 1}, \bC_{{\sf D},k})$ is a DL small-scale fading vector.
We assume that $\bg_{{\sf CCI},k} = \sqrt{\rho_{{\sf CCI},k}} \tilde{\bg}_{{\sf CCI},k}$, where $\rho_{{\sf CCI},k}$ is the CCI channel path-loss and $\tilde{\bg}_{{\sf CCI},k} \! \sim \! \cC\cN(\mathbf{0}_{K_{\sf U} \times 1}, \bI_{K_{\sf U}})$.

%%%%% SQNR Analysis %%%%%
% SQNR 기반의 분석 --> quantization bits에 따른 MRT & ZF의 SQNR 성능 분석
\section{Per-Antenna SQNR Analysis}
\label{sec:ADC}
Recent studies on HD systems have shown that employing low-resolution quantizers can significantly reduce the power consumption while achieving comparable SE to the system with high-resolution quantizers by effectively managing the quantization errors \cite{choi:tsp:2017, choi:twc:2022, dai:commlett:2021, hu:tcom:2019, kim:iotj:2022, park:twc:23}.
In FD systems, however, the strong residual analog SI poses a major challenge to employing low-resolution ADCs as discussed in Remark~\ref{rm:remark1}, underscoring the critical interplay between the quantization effects and the SI.
Accordingly, we aim to characterize the impact of employing the low-resolution ADCs and analyze the desirable ADC resolution for the predefined system requirement. 

To this end, we assume high-resolution DACs to focus solely on the effect of low-resolution ADCs.
We also assume an interference-limited regime, since the residual analog SI and the UL signal typically far dominate the additive thermal noise.
In addition, we assume that the ADC input always falls within the ADC dynamic range by considering perfect automatic gain control.
In this section, we denote $b \! = \! b_{\sf ADC}$, $\alpha \! = \! \alpha_{\sf ADC}$, and $\beta \! = \! \beta_{\sf ADC}$ for notational simplicity, and assume $\alpha = 1-\frac{\pi \sqrt3}{2}2^{-2b}$ for analytical tractability unless mentioned otherwise.

We define the UL signal-to-quantization-noise ratio (SQNR) for the $n$th receive antenna of the AP and the $i$th UL user based on \eqref{eq:r_ac_Uq} and \eqref{eq:R_qADCqADC} with proper simplification as
\begin{align}
    \nonumber
    {\rm SQNR}_{n,i}({\bW},b) 
    &= \frac{\mathbb{E}\left[\left|\boldsymbol{\Phi}_{\alpha_{\sf ADC}, \left(n, n\right)} \bH_{{\sf U}, \left(n, i\right)} s_{{\sf  U}, i}\right|^{2}\right]
    }{
    \left[\bR_{\bq_{\sf ADC}\bq_{\sf ADC}}(\bW)\right]_{\left(n,n\right)}
    }
    \\\label{eq:SQNR}
    &= \! \frac{ \alpha \left|\bH_{{\sf U}, \left(n, i\right)} \right|^{2} }{\beta \! \left(\sum_{k=1}^{K_{\sf U}} \left|\bH_{{\sf U}, \left(n, k\right)} \right|^{2} \! + \! \frac{P_{\sf D}}{P_{\sf U}} \|{\bg}^{\sf H}_{{\sf SI}, n} {\bW} \|^{2} \right)}.
\end{align}
We note that the per-antenna SQNR in \eqref{eq:SQNR} indicates the power ratio between the desired UL signal and quantization noise coupled with the residual analog SI.
Consequently, by analyzing the SQNR, the interplay between quantization effects and SI can be properly characterized as a function of system parameters such as the number of quantization bits, transmit power, path-loss, and analog SI capability.

Assuming that there exists a required average SQNR for reliable FD communications, we formulate the following problem to determine the required number of ADC bits for achieving such requirement:
\begin{align}
    \label{eq:prob_sqnr}
    \underset{b}{\text{minimize}}&\;\; b
    \\
    \label{eq:constraint_prob_sqnr}
    \text{subject to}&\;\; \mathbb{E}[{\rm SQNR}_{n,i} ({\bf W},b)]  
    \geq  \tau \quad \forall n,i,
\end{align}
where $\tau$ is the target average SQNR threshold.
In the following subsections, we derive the minimum required number of bits for $(i)$ SI-independent linear precoders and $(ii)$ SI-dependent linear precoder by solving the problem in \eqref{eq:prob_sqnr}.

%%%%%%%%%%%
\subsection{SI-Independent Precoder}
We first derive a lower bound on the average SQNR for the SI-independent precoding.
For instance, such linear precoders include MRT and ZF without nullifying the SI channel.
\begin{theorem}[Average SQNR lower bound]
\label{Th:theorem2}
    For linear precoders that are statistically independent of the SI channels, the lower bound on the average SQNR in \eqref{eq:SQNR} is derived as
    \begin{align}
        \mathbb{E}[{\rm SQNR}_{n,i} ({\bW},b)] \geq  
        \frac{\alpha}{1-\alpha} \left( 1 - \gamma_{i}e^{\gamma_{i}} \Gamma(0, \gamma_{i}) \right), \label{eq:SQNR_LB_MRT}
    \end{align}
    where $\gamma_{i}=\frac{\sum_{k \neq i, k=1}^{K_{\sf U}} P_{\sf U} \rho_{{\sf U}, k} + P_{\sf D} \kappa_{\sf a}}{P_{\sf U} \rho_{{\sf U},i}}$, and $\Gamma(a,z)=\int_z^{\infty}t^{a-1}e^{-t} \mathrm{d}t$ is the upper incomplete gamma function. 
    \begin{proof}
    Let $X_{k}=P_{\sf U}\left| {\bf H}_{{\sf U}, \left(n, k \right)} \right|^2$ and $Y = P_{\sf D}\| \bg_{{\sf SI}, n}^{\sf H} \bW \|^2$. 
    Then, the average SQNR is simply represented as
    \begin{align}
        \label{eq:MU_Simple_SQNR_LB}
        \mathbb{E}[{\rm SQNR}_{n, i} ({\bW},b)] = \frac{\alpha}{1-\alpha} \mathbb{E} \left[ \frac{X_{i}}{\sum_{k=1}^{K_{\sf U}} X_{k} + Y}  \right],
    \end{align}
    where each $X_k$ is an exponential random variable with mean $P_{\sf U}\rho_{{\sf U}, k}$.
    Leveraging the independence between the considered class of linear precoders and the SI channel, and applying the IID property of the SI channel, we first derive $ \mathbb{E} \left[Y \right]$ as
    \begin{align}
        \mathbb{E} \left[Y \right] &= P_{\sf D}\mathbb{E} \left[ \| {\bg}^{\sf H}_{{\sf SI}, n} {\bW} \|^2 \right] \nonumber \\
        &= P_{\sf D} \mathbb{E}_{\bW} \left[ \mathbb{E}_{{\bg}_{{\sf SI}, n}} \left[ {\bg}^{\sf H}_{{\sf SI}, n} {\bW} {\bW}^{\sf H} {\bg}_{{\sf SI}, n}\bigm|{\bW} \right] \right]\nonumber \\
        &= P_{\sf D} \kappa_{\sf a}\mathbb{E}_{\bW} \left[ \mathrm{Tr}\left({\bW}{\bW}^{\sf H}\right)\right] 
        \stackrel{(a)}= P_{\sf D} \kappa_{\sf a}, \label{eq:Expectation_Y}
    \end{align}
    where $(a)$ follows from that the precoder is normalized due to the transmit power constraint.
    Since all $X_k$ and $Y$ are independent, and \eqref{eq:MU_Simple_SQNR_LB} is a convex function of $Y$ and all $X_k$ except $X_{i}$, we apply Jensen's inequality to each random variable individually, except $X_i$, and use $\mathbb{E}[X_k] = P_{\sf U} \rho_{{\sf U},k}$ and  $\mathbb{E}[Y] = P_{\sf D} \kappa_{\sf a}$ in \eqref{eq:Expectation_Y} to derive the following bound:
    % 수정한 부분 확인했습니다. - 승형, 재현
    \begin{align}
    &\mathbb{E}[{\rm SQNR}_{n,i} ({\bW}, b)] \! \geq \! \frac{\alpha}{1-\alpha} \mathbb{E}\left[ \! \frac{X_i}{X_i \! + \! \sum_{k=1, k \neq i}^{K_{\sf U}} P_{\sf U} \rho_{{\sf U},k} \! + \! P_{\sf D} \kappa_{\sf a}} \! \right] \nonumber \\
    &\stackrel{(b)}= \! \frac{\alpha}{1 \! - \! \alpha} \!\! \int_{0}^{\infty} \frac{P_{\sf U}\rho_{{\sf U}, i}t e^{-t}}{P_{\sf U}\rho_{{\sf U}, i}t+\sum_{k = 1, k \neq i}^{K_{\sf U}}P_{\sf U}\rho_{{\sf U}, k} + P_{\sf D}\kappa_{\sf a}} \mathrm{d}t \nonumber \\
    &= \! \frac{\alpha}{1 \! - \! \alpha} \!\! \int_{0}^{\infty} \!\! \left(1 \! -\!  \frac{\gamma_{i}}{t+\gamma_{i}}\right) \! e^{-t} \mathrm{d}t \!=\! \frac{\alpha}{1-\alpha} \! \left( \! 1 \!- \! \gamma_{i}e^{\gamma_{i}} \Gamma(0, \gamma_{i}) \right),
    \end{align}
    where $(b)$ comes from $|{\bf H}_{{\sf U},(n,i)}|^2\sim {\rm exp}(\frac{1}{\rho_{{\sf U}, i}})$.
    This completes the proof.
    \end{proof}
\end{theorem}
 We note that since $\mathrm{Tr}({\bW}{\bW}^{\sf H})=1$ is always enforced due to the transmit power constraint, 
% we obtain $\mathbb{E}[P_{\sf D}\|{\bg}^{\sf H}_{{\sf SI}, n}{\bf W}\|^2] \! = \! P_{\sf D}\kappa_{\sf a}$. 
the lower bound in \eqref{eq:SQNR_LB_MRT} holds for any linear precoder that is independent to the SI channels.
Using \eqref{eq:SQNR_LB_MRT}, we  derive the following corollary of the minimum required ADC bits satisfying constraint in \eqref{eq:constraint_prob_sqnr}.

\begin{corollary}[Upper bound of minimum number of bits]
    \label{cor:b_ind_UB}
   The minimum number of ADC bits for any linear precoder independent of the UL and SI channels is upper bounded by
    \begin{align}
        \label{eq:b_Arbitrary_LB}
        b^{\sf Ind,UB} \! = \! \max_{1 \leq i \leq K_{\sf U}} \frac{1}{2} \log_{2} \! \left( \frac{\pi \sqrt3}{2} \left(1+ \frac{\tau}{1-\gamma_{i}e^{\gamma_{i}}\Gamma(0,\gamma_{i})}\right)  \right),
    \end{align}
    where $\gamma_{i}=\frac{\sum_{k \neq i, k=1}^{K_{\sf U}} P_{\sf U} \rho_{{\sf U}, k} + P_{\sf D} \kappa_{\sf a}}{P_{\sf U} \rho_{{\sf U},i}}$.
    \begin{proof}
    By substituting the lower bound \eqref{eq:SQNR_LB_MRT} into \eqref{eq:constraint_prob_sqnr}, 
    % and setting $\alpha = 1-\frac{\pi \sqrt3}{2}2^{-2b}$, 
    we relax the problem in \eqref{eq:prob_sqnr}.
    Therefore, the minimum required ADC bits of the relaxed problem for an arbitrary precoder are given by \eqref{eq:b_Arbitrary_LB}, which is the upper bound on the minimum required number of bits.
    This completes the proof.
    \end{proof}
\end{corollary}
\begin{corollary}
    \label{cor:b_LB_approx}
    For large $\gamma_{i}$, \eqref{eq:b_Arbitrary_LB} is approximated as
    \begin{align}
         \label{eq:b_ind_aUB}
         b^{\sf Ind,aUB} \approx \max_{1 \leq i \leq K_{\sf U}} \frac{1}{2} \log_{2} \left( \frac{\pi \sqrt3}{2} \left(1+ \tau \gamma_{i} \right) \right).
    \end{align}
    \begin{proof}
        Based on \cite{abramowitz1965handbook}, the lower bound on the average SQNR in Theorem~\ref{Th:theorem2} is approximated for large $\gamma_{i}$ as
    \begin{align}  
        \frac{\alpha}{1-\alpha} \left( 1 - \gamma_{i}e^{\gamma_{i}} \Gamma(0, \gamma_{i}) \right) \approx \frac{\alpha}{(1-\alpha)\gamma_{i}}.
    \end{align}
    Using this approximation, we follow the proof of Corollary~\ref{cor:b_ind_UB}. 
    This completes the proof.
    \end{proof}
\end{corollary}
It is shown from Corollary~\ref{cor:b_LB_approx} that the approximated upper bound on the number of ADC bits increases logarithmically with respect to the ratio of the inter-user interference plus the residual SI power over the desired UL user power $\gamma_i$.
In addition, such approximation in \eqref{eq:b_ind_aUB} is considered to be tight since the residual analog SI power $P_{\sf D}\kappa_{\sf a}$ dominates the UL received signal $P_{\sf U}\rho_{{\sf U},i}$ in FD systems.

As a special case, when there exists a single DL and UL user, we derive a exact closed-form expression of the average SQNR with the MRT precoder.

\begin{theorem}[Special case: average SQNR in single-user case]
\label{Th:theorem1}
    Let $K_{\sf U}=K_{\sf D}=1$. 
    For the $n$th receive antenna of the AP and a single UL user, the  average SQNR  in \eqref{eq:SQNR}  is derived as
    \begin{align}
        \label{eq:SU_MRT_Exact_SQNR}
        \mathbb{E}[{\rm SQNR}_{n, 1} ({\bw}^{\sf MRT},b)] = \frac{\alpha}{1-\alpha} \left[ \frac{\gamma \ln \gamma}{\left( \gamma - 1  \right)^{2}} - \frac{1}{\gamma - 1} \right],
    \end{align}
    where ${\bw}^{\sf MRT}$ is the MRT precoder and $\gamma=\frac{P_{\sf D}\kappa_{\sf a}}{P_{\sf U}\rho_{{\sf U}, 1}}$ is the ratio of the residual analog SI power and received UL signal power.
    \begin{proof}
    Putting $K_{\sf U} \! = \! K_{\sf D} \! = \! 1$ and ${\bW} = {\bw}^{\sf MRT} = {\bh_{{\sf D},1}}/{\|\bh_{{\sf D},1} \|}$ in \eqref{eq:SQNR}, the average SQNR reduces to
    \begin{align}
        & \bbE[{\rm SQNR}_{n, 1} ({\bw}^{\sf MRT},b)] = \frac{\alpha}{1-\alpha} \bbE \left[ \frac{X}{X + Y}  \right], \label{eq:SU_Simple_SQNR}
    \end{align}
    where $X \!\!=\! P_{\sf U}| {\bf H}_{{\sf U}, (n,1)} |^2$ and $Y \!\!=\! P_{\sf D} \|{\bg}^{\sf H}_{{\sf SI}, n}{\bw}^{\sf MRT}\|^2$.
    The random variables $X$ and $Y$ are mutually independent and each follows an exponential distribution with mean $P_{\sf U}\rho_{{\sf U}, 1}$ and $P_{\sf D}\kappa_{\sf a}$, respectively. 
    To derive the average SQNR in \eqref{eq:SU_Simple_SQNR}, we calculate the cumulative density function (CDF) of $Z= X/Y$ as
    \begin{align}
        F_{Z}\left(z \right) &=\! \mathbb{P} \! \left( \frac{X}{Y} \! \leq \! z \right) \! = \!\!\! \int_{0}^{\infty} \!\! \int_{0}^{yz} \!\! \frac{1}{P_{\sf U}\rho_{{\sf U}, 1}P_{\sf D}\kappa_{\sf a}} e^{-\left( \frac{x}{P_{\sf U}\rho_{{\sf U}, 1}} + \frac{y}{P_{\sf D}\kappa_{\sf a}}\right)}\mathrm{d}x \mathrm{d}y \nonumber \\
        &= \! \frac{P_{\sf D} \kappa_{\sf a} z}{P_{\sf D} \kappa_{\sf a} z + P_{\sf U}\rho_{{\sf U}, 1}}. \label{eq:CDF_Z}
    \end{align}
    Based on the Fubini theorem \cite{veraar2012stochastic} and non-negativity of $\frac{X}{X+Y}$, $\bbE[\frac{X}{X+Y}]$ in \eqref{eq:SU_Simple_SQNR} is given by 
    \begin{align}
        \label{eq:expect_Theo1}
        \mathbb{E} \left[ \frac{X}{X + Y}  \right] = \int_0^{\infty} \mathbb{P}\left(\frac{X}{X+Y} > t \right) \mathrm{d}t.
    \end{align}
    Using \eqref{eq:CDF_Z}, we calculate an integral in \eqref{eq:expect_Theo1} as
    \begin{align}
        &\int_0^{\infty} \mathbb{P}\left(\frac{X}{X+Y} > t \right) \mathrm{d}t = \int_0^1 \mathbb{P}\left(\frac{X}{Y} > \frac{t}{1-t} \right) \mathrm{d}t \nonumber \\
        &= \frac{P_{\sf U}\rho_{{\sf U}, 1}P_{\sf D}\kappa_{\sf a}}{\left( P_{\sf U}\rho_{{\sf U}, 1} - P_{\sf D}\kappa_{\sf a} \right)^{2}} \ln{\frac{P_{\sf D}\kappa_{\sf a}}{P_{\sf U}\rho_{{\sf U}, 1}}} \! - \! \frac{P_{\sf U}\rho_{{\sf U}, 1}}{P_{\sf D}\kappa_{\sf a}-P_{\sf U}\rho_{{\sf U}, 1}}. \label{eq:tail_prob}
    \end{align}
    Substituting \eqref{eq:tail_prob} into \eqref{eq:SU_Simple_SQNR} yields the closed-form average SQNR in~\eqref{eq:SU_MRT_Exact_SQNR}. 
    This completes the proof.
    \end{proof}
\end{theorem}

Based on Theorem~\ref{Th:theorem1}, we derive the following corollary.
\begin{corollary}[Special case: minimum number of bits in single-user case]
    \label{cor:b_MRT_SU}
    When using the MRT precoder with $K_{\sf U} \! = \!  K_{\sf D} \! = \! 1$, the minimum required ADC bits to satisfy \eqref{eq:constraint_prob_sqnr} are given by
    \begin{align}
        \label{eq:SU_b_MRT}
        b^{\sf MRT} =  \frac{1}{2} \log_2\left(\frac{\pi\sqrt3}{2}\left(1+\frac{\tau \left( \gamma-1 \right)^2}{\gamma \ln{\gamma} - \gamma + 1}\right) \right).
    \end{align}
    \begin{proof}
    Putting $\alpha \! = \!\! 1 - \frac{\pi \sqrt3}{2}2^{-2b}$ into \eqref{eq:SU_MRT_Exact_SQNR}, the minimum required number of ADC bits satisfying the constraint \eqref{eq:constraint_prob_sqnr} are derived as \eqref{eq:SU_b_MRT}. This completes the proof.
    \end{proof}
\end{corollary}
Similar to \eqref{eq:b_ind_aUB}, the exact minimum required number of ADC bits with single-user MRT precoding in Corollary~\ref{cor:b_MRT_SU} also increases with the ratio of the residual analog SI power over the received UL signal power $\gamma$, which increases logarithmically for large $\gamma$.
Accordingly, we have the following remark.
\begin{remark}
    \normalfont The derived results in the corollaries theoretically indicate that unlike the HD systems, the residual analog SI significantly prevents the FD systems from lowering the ADC resolution.
\end{remark}

The considered precoders represent the worst case in terms of reducing the required ADC resolution since they ignore the SI channel.
In the following subsection, we analyze the required number of ADC bits when the precoder fully nullifies the SI channel, which corresponds to the best case analysis.

\subsection{SI-Dependent Precoder: ZF-NSI}
We assume that the ZF precoder is designed to nullify the SI channel (NSI) to the $n$th AP's receive antenna as well as the DL channels, and hence it is a SI-dependent precoder.
We also assume distinct path-losses among all UL users, i.e., $\rho_{{\sf U},i} \neq \rho_{{\sf U},j}$, $\forall i\neq j$ with $K_{\sf U}>1$.
\begin{theorem}[Average SQNR]
    When the SI to the $n$th receive antenna at the AP is fully nullified by the ZF precoder (ZF-NSI), the average SQNR  in \eqref{eq:SQNR} is derived as
    \begin{align}
        \label{eq:ZF_Exact_SQNR}
        &\mathbb{E}[{\rm SQNR}_{n,i} ({\bW}^{\sf ZF-NSI},b)] = \frac{\alpha}{1-\alpha} \rho_{{\sf U}, i} \\
        &\times \! \sum_{k=1, k \neq i}^{K_{\sf U}} \Bigg[ \! \bigg( \prod_{j=1, j \notin \{i,k\}}^{K_{\sf U}} \! \frac{\rho_{{\sf U},k}}{\bar{\rho}_{{\sf U},k,j}} \bigg) \! \left( \frac{ \rho_{{\sf U}, k} \ln\left( {\rho_{{\sf U}, k}}/{\rho_{{\sf U}, i}} \right) }{( \bar{\rho}_{{\sf U},k,i} )^2} - \frac{1}{\bar{\rho}_{{\sf U},k,i}} \right) \! \Bigg], \nonumber
    \end{align}
    where ${\bW}^{\sf ZF-NSI}$ is the ZF-NSI precoder,
    $\bar{\rho}_{{\sf U},k,j} = \rho_{{\sf U},k} - \rho_{{\sf U},j}$, and $\bar{\rho}_{{\sf U},k,i} = \rho_{{\sf U},k} - \rho_{{\sf U},i}$.
    \begin{proof}
    When the ZF-NSI precoder is employed, the SQNR is given by \eqref{eq:SQNR} with ${\bg}^{\sf H}_{{\sf SI}, i} {\bW}^{\sf ZF-NSI} = {\bf 0}_{1 \times K_{\sf D}}$. 
    Using the same notation in \eqref{eq:MU_Simple_SQNR_LB}, the average SQNR is expressed as 
    \begin{align}
        \label{eq:ZF_Exact_Ergodic}
        \mathbb{E}[{\rm SQNR}_{n,i} ({\bW}^{\sf ZF-NSI},b)] \!=\! \frac{\alpha}{1-\alpha} \mathbb{E} \! \left[ \! \frac{X_i}{X_i + \sum_{k=1, k \ne i}^{K_{\sf U}}X_k} \! \right] \!.
    \end{align}
    The random variable $X_k$ follows an independent exponential distribution with mean $P_{\sf U}\rho_{{\sf U}, k}$. 
    For deriving a closed-form average SQNR, we set $Z_{i} = \sum_{k=1, k \ne i}^{K_{\sf U}}X_k$.
    The random variable $Z_{i}$ follows a hypoexponential distribution with the following probability density function (PDF) in \cite{ross2014introduction}, i.e.,
    \begin{align}
        f_{Z_{i}}(z) =
         \sum_{k=1, k \neq i}^{K_{\sf U}} \frac{C_{k, i}}{P_{\sf U} \rho_{{\sf U}, k}}e^{- \frac{z}{P_{\sf U}\rho_{{\sf U}, k}}},
    \end{align}
    where $C_{k, i} = \prod_{j=1, j \notin \{i,k\}}^{K_{\sf U}}\frac{\rho_{{\sf U},k}}{\rho_{{\sf U}, k}-\rho_{{\sf U}, j}}$.
    Subsequently, the expectation in \eqref{eq:ZF_Exact_Ergodic} is calculated as  
    \begin{align}
        &\mathbb{E} \left[ \frac{X_i}{X_i + \sum_{k=1, k \ne i}^{K_{\sf U}}X_k} \right] = \int_{0}^{\infty} \int_{0}^{\infty} \frac{x}{x + z} f_{X_{i}}(x)f_{Z_{i}}(z) \mathrm{d}z \mathrm{d}x \nonumber \\
        &\underset{=}{(a)} \sum_{k=1, k \neq i}^{K_{\sf U}} C_{k, i}
        \int_{0}^{\infty} \int_{0}^{\infty} 
        \frac{x}{x + z}  \frac{e^{-\frac{x}{P_{\sf U} \rho_{{\sf U}, i}}} }{P_{\sf U} \rho_{{\sf U}, i}} \frac{e^{-\frac{z}{P_{\sf U} \rho_{{\sf U}, k}}} }{P_{\sf U} \rho_{{\sf U}, k}}
        \mathrm{d}z \mathrm{d}x \nonumber \\
        &= \rho_{{\sf U}, i}\sum_{k=1, k \neq i}^{K_{\sf U}} C_{k, i}
        \Bigg( \frac{ \rho_{{\sf U}, k} \ln ( \rho_{{\sf U}, k}/\rho_{{\sf U}, i} ) }{( \rho_{{\sf U}, k} - \rho_{{\sf U}, i} )^2} - \frac{1}{\rho_{{\sf U}, k} - \rho_{{\sf U}, i}} \Bigg)
        \nonumber \\ \label{eq:G}
        &\triangleq G_i.
    \end{align}
    Here, $(a)$ is obtained by applying the result from Theorem~\ref{Th:theorem1}.
    This is valid because the integral represents the expected value of a function of two independent exponential random variables with different means.
    This completes the proof.
    \end{proof}
\end{theorem}
Based on \eqref{eq:ZF_Exact_SQNR}, we derive the following corollary.

\begin{corollary}[Minimum number of bits]
    \label{cor:b_ZF_NSI}
    Using the ZF-NSI precoder, the minimum required ADC bits are given by
    \begin{align}
        \label{eq:b_ZF_SIC}
        b^{\sf ZF-NSI} = \max_{1 \leq i \leq K_{\sf U}} \frac{1}{2} \log_2\bigg(\frac{\pi\sqrt3}{2}\left(1+\frac{\tau}{G_{i}}\right) \bigg),
    \end{align} 
    where $G_i$ is defined in \eqref{eq:G}.
    \begin{proof}
    % 수정된 부분 확인했습니다. - 승형, 재현
    By substituting $\alpha \! = \!\! 1 \!\! - \! \frac{\pi \sqrt3}{2}2^{-2b}$ into \eqref{eq:ZF_Exact_SQNR}, as in Corollary~\ref{cor:b_MRT_SU}, the solution of the problem in \eqref{eq:prob_sqnr} becomes \eqref{eq:b_ZF_SIC}. 
    This completes the proof.   
    \end{proof}
\end{corollary}
As the ZF-NSI precoder nullifies the SI channel, \eqref{eq:b_ZF_SIC} is no longer a function of the residual SI, i.e., the ZF-NSI precoder perfectly eliminates the SI, lowering the quantization noise floor, thereby reducing the required number of ADC bits.
Conversely, the MRT precoder or other precoders independent of the SI channels aim to maximize the DL SE while ignoring the impact on the SI, which requires more ADC bits to meet the target UL SQNR.
Therefore, an optimal precoder that jointly balances SI reduction and DL signal amplification achieves a superior trade-off and higher sum SE.
Then, assuming that there exists an optimal precoder ${\bf W}^\star$ that maximizes the sum SE of the considered FD system with the SQNR constraint, we naturally establish the following proposition.

\begin{proposition}
    \label{prop:proposition1}
    The minimum required number of bits for the optimal precoder ${\bf W}^\star$ is bounded by
    \begin{align}
         b^{\sf ZF-NSI} \leq b^{\star} \leq b^{\sf Ind, UB}.
    \end{align}
\end{proposition}

Now, to provide a more tractable analysis, we consider a homogeneous path-loss scenario where all UL users have the same path-loss, and derive the following corollary.
\begin{corollary}[Special case: homogeneous path-loss]
As a special case, we consider $\rho_{{\sf U},k}=\rho_{\sf U}$ for $1\le k\le K_{\sf U}$. 
Then, the minimum required number of ADC bits in \eqref{eq:b_Arbitrary_LB} and \eqref{eq:b_ZF_SIC} are respectively represented as 
\begin{align}
\label{eq:bit_symuser_noSIC}
\bar{b}^{\sf Ind, UB} &= \frac{1}{2} \log_{2} \bigg( \frac{\pi \sqrt3}{2} \left(1 + \frac{\tau}{ 1 - \eta e^{\eta} \Gamma(0, \eta)} \right) \bigg), \\
\label{eq:bit_symuser_SIC}
\bar{b}^{\sf ZF-NSI} &=\frac{1}{2}\log_{2}\bigg( \frac{\pi \sqrt3}{2} \left(1+ \tau K_{\sf U} \right) \bigg), 
\end{align}
where $\eta = K_{\sf U} - 1 + \frac{ P_{\sf D} \kappa_{\sf a}}{P_{\sf U} \rho_{\sf U}}$. 
\begin{proof}
    By substituting $\rho_{{\sf U}, k} = \rho_{\sf U}$ for $1 \le k \le K_{\sf U}$ in \eqref{eq:b_Arbitrary_LB}, the upper bound on the minimum required number of bits in \eqref{eq:b_Arbitrary_LB} reduces to \eqref{eq:bit_symuser_noSIC}. 
    For SI-dependent precoder, \eqref{eq:ZF_Exact_Ergodic} reduces to $\frac{\alpha}{1-\alpha} \cdot \frac{1}{K_{\sf U}}$ since each $X_{k}$ follows an exponential distribution with the same mean $P_{\sf U}\rho_{\sf U}$.
    Similar to Corollary~\ref{cor:b_MRT_SU} and Corollary~\ref{cor:b_ZF_NSI}, by putting $\alpha \! = \!\! 1 \!\! - \! \frac{\pi \sqrt3}{2}2^{-2b}$ into \eqref{eq:ZF_Exact_Ergodic}, the minimum required number of  bits becomes \eqref{eq:bit_symuser_SIC}.
    This completes the proof.
\end{proof}
\end{corollary}
From Proposition~\ref{prop:proposition1}, \eqref{eq:bit_symuser_noSIC}, and \eqref{eq:bit_symuser_SIC}, we derive the gap between the minimum required number of bits with linear precoders for the homogeneous path-loss scenario.

\begin{proposition}
\label{prop:proposition2}
Let $\rho_{{\sf U},k} \! = \! \rho_{\sf U}$ for $1 \! \le \! k \! \le \! K_{\sf U}$. 
The upper bound on the gap between the minimum required ADC bits in \eqref{eq:b_Arbitrary_LB} and \eqref{eq:b_ZF_SIC} is approximated for large $\eta$ as
\begin{align}
    \label{eq:Dbmax_approx}
    \Delta \bar{b}_{\sf max}
    \approx \frac{1}{2}\log_{2} \left( \frac{1+\tau \eta}{1+\tau K_{\sf U}} \right).
\end{align}
\begin{proof}
    For large $\eta$, the required ADC bit in \eqref{eq:bit_symuser_noSIC} is approximated as $\bar{b}^{\sf Ind, UB} \approx \frac{1}{2} \log_{2} ( \frac{\pi \sqrt{3}}{2} (1 + \tau \eta) )$.
    Then we directly compute 
    $\Delta \bar{b}_{\sf max} = \bar{b}^{\sf Ind, UB} - \bar{b}^{\sf ZF-NSI}$ with the approximation.
\end{proof}
\end{proposition}
Proposition~\ref{prop:proposition2} measures the upper bound on how many bits can be reduced by fully nullifying the SI channel.
From \eqref{eq:Dbmax_approx}, it is observed that the required number of bits increases logarithmically with respect to the residual analog SI power $P_{\sf D}\kappa_{\sf a}$ for guaranteeing the target SQNR $\tau$.
In the following sections, we optimize the FD beamforming to maximize the sum SE and numerically confirm that the number of ADC bits corresponding to the proposed precoder also satisfies \eqref{eq:Dbmax_approx} as the precoder well balances the SI nullification for the UL SE and DL SINR maximization for the DL SE.

%%%%%%%%%%%% propose the precoding method %%%%%%%%%%%%
\section{Proposed FD Joint Beamforming Algorithm}
\label{sec:beamforming}
We propose an advanced FD MU-MISO beamforming algorithm that jointly considers residual SI, IUI, CCI, and quantization errors for given ADC and DAC configurations with the goal of maximizing the UL and DL sum SE.

%%%%%%%%%%%%%%%
\subsection{Performance Metric and Problem Formulation}
For the FD MU-MISO system described in Section~\ref{sec:system}, the closed-form SE expressions of the $k$th DL and UL users are respectively characterized as
\begin{align}
    \label{eq:R_Dk}
    &R_{{\sf D},k}({\bf W}) \! = \! \mathrm{log}_2 \! \left(1 \! + \! \frac{P_{\sf D}|\bh_{{\sf D},k}^{\sf H} \boldsymbol{\Phi}_{\alpha_{\sf DAC}} \bw_{k}|^{2}}{\mathrm{IUI}_{{\sf D},k} + \mathrm{QN}_{{\sf DAC},k} + I_{{\sf CCI},k} + \sigma_{\sf D}^2} \right),\\
    &R_{{\sf U},k}({\bf W},{\bf F}) \nonumber\\
    \label{eq:R_Uk}
    &= \! \mathrm{log}_2 \left( \! 1 + \frac{P_{\sf U}|\bff_{k}^{\sf H} \boldsymbol{\Phi}_{\alpha_{\sf ADC}} \bh_{{\sf U},k}|^{2}}{\mathrm{IUI}_{{\sf U},k} \! + \! \mathrm{QN}_{{\sf ADC},k} \! + \! I_{{\sf SI},k} \! + \!  \|\bff_{k}^{\sf H} \boldsymbol{\Phi}_{\alpha_{\sf ADC}}\|^2 \sigma_{\sf U}^2} \! \right),
\end{align}
where
\begin{align}
    \mathrm{IUI}_{{\sf D},k} =&\, P_{\sf D} \sum_{i=1, i \ne k}^{K_{\sf D}} |\bh_{{\sf D},k}^{\sf H} \boldsymbol{\Phi}_{\alpha_{\sf DAC}} \bw_i|^2, 
    \\
    \mathrm{IUI}_{{\sf U},k} =&\, P_{\sf U}\sum_{i=1, i \ne k}^{K_{\sf U}} |\bff_k^{\sf H} \boldsymbol{\Phi}_{\alpha_{\sf ADC}} \bh_{{\sf U},i}|^2, 
\end{align}
\begin{align}
    {I}_{{\sf CCI},k} =&\, P_{\sf U}\|\bg_{{\sf CCI},k}\|^{2},\label{eq:I_CCI}
    \\
    I_{{\sf SI},k} =&\, P_{\sf D} \kappa_{\sf d} \sum_{i=1}^{K_{\sf D}} |\bff_{k}^{\sf H} \boldsymbol{\Phi}_{\alpha_{\sf ADC}} \bG^{\sf H}_{\sf SI} \boldsymbol{\Phi}_{\alpha_{\sf DAC}} \bw_{i}|^2 \label{eq:I_SI}
    \\
    &+\kappa_{\sf d} \bff_{k}^{\sf H} \boldsymbol{\Phi}_{\alpha_{\sf ADC}} \bG^{\sf H}_{\sf SI} \bR_{\bq_{\sf {DAC}}\bq_{\sf {DAC}}}({\bf W}) \bG_{\sf SI} \boldsymbol{\Phi}_{\alpha_{\sf ADC}}^{\sf H} \bff_{k}, \nonumber
    \\
    \mathrm{QN}_{{\sf DAC},k} =&\, \bh_{{\sf D},k}^{\sf H} \bR_{\bq_{\sf {DAC}}\bq_{\sf {DAC}}}({\bf W}) \bh_{{\sf D},k}, \label{eq:QN_DAC}
    \\
     \mathrm{QN}_{{\sf ADC},k} =&\, \bff_{k}^{\sf H} \bR_{\bq_{\sf {ADC}}\bq_{\sf {ADC}}}({\bf W}) \bff_{k} 
     \label{eq:QN_ADC}.
\end{align}
In addition, the DL transmit power constraint is computed as
\begin{align}
    \label{eq:expected_X_Dq}
    &\mathrm{Tr} \left(\bbE \left[\bx_{\sf D,q} \bx_{\sf D,q}^{\sf H} \right]\right) \nonumber \\
    &= \mathrm{Tr} \bigg( P_{\sf D} \boldsymbol{\Phi}_{\alpha_{\sf DAC}} \bW \bW^{\sf H}\boldsymbol{\Phi}_{\alpha_{\sf DAC}}^{\sf H} + \boldsymbol{\Phi}_{\alpha_{\sf DAC}} \boldsymbol{\Phi}_{\beta_{\sf DAC}} \mathrm{diag}{\left\{ P_{\sf D} \bW \bW^{\sf H} \right\}} \bigg) \nonumber\\
    &\overset{(a)}{=} \mathrm{Tr} \left( P_{\sf D} \boldsymbol{\Phi}_{\alpha_{\sf DAC}} \bW \bW^{\sf H} \right), 
\end{align}
where $(a)$ is derived from the relationship between $\boldsymbol{\Phi}_{\alpha_{\sf DAC}}$ and $\boldsymbol{\Phi}_{\beta_{\sf DAC}}$, i.e., $\boldsymbol{\Phi}_{\alpha_{\sf DAC}} + \boldsymbol{\Phi}_{\beta_{\sf DAC}} = \bI_{N_t}$, and the trace property.

Thus, we formulate a sum SE maximization problem as
\begin{align}
    \underset{\bW, \bF}{\mathrm{maximize}}& \;\; \sum_{k=1}^{K_{\sf D}} R_{{\sf D},k}(\bW) + \sum_{k=1}^{K_{\sf U}} R_{{\sf U},k}(\bW;\bF) \label{eq:objective}\\
    \mathrm{subject \; to}& \;\; \mathrm{Tr} \left( \boldsymbol{\Phi}_{\alpha_{\sf DAC}} \bW \bW^{\sf H} \right) 
    \leq  1,
    \label{eq:constraint}
\end{align}
where \eqref{eq:constraint} is a transmit power constraint at the AP.
Perfect channel state information (CSI) for both UL and DL is assumed to be available at the AP.
We note that solving the optimization problem \eqref{eq:objective} is particularly challenging because the residual SI, CCI, and quantization errors are tightly intertwined with the precoder and combiner, significantly complicating the optimization process.
To address these challenges, we first propose an advanced precoder optimization method.

%%%%% Rayleigh reformulation %%%%%
\subsection{Beamforming Optimization}
Now, we first consider the precoder optimization subproblem of \eqref{eq:objective} with a fixed combiner.
Even with a fixed combiner, the problem in \eqref{eq:objective} remains non-convex, making it infeasible to find a global optimal solution.
Therefore, we convert the problem into a tractable form and derive a first-order optimality condition to identify local optimal solutions.

%%%%%%%%%%%
\subsubsection{Problem Reformulation}
We rewrite the problem into a tractable form to utilize the GPI method \cite{choi:twc:19}; the DAC quantization noise-related term in \eqref{eq:QN_DAC} is reformulated as
\begin{align}
    {\mathrm{QN}_{{\sf DAC},k}} &= P_{\sf D}\bh_{{\sf D},k}^{\sf H}\boldsymbol{\Phi}_{\alpha_{\sf DAC}} \boldsymbol{\Phi}_{\beta_{\sf DAC}} \mathrm{diag}{\left\{ \sum_{i=1}^{K_{\sf D}}\bw_i \bw^{\sf H}_i \! \right\}} \bh_{{\sf D},k} \nonumber\\
    &= P_{\sf D}\sum_{i=1}^{K_{\sf D}} \! \bw^{\sf H}_{i} \boldsymbol{\Phi}_{\alpha_{\sf DAC}} \boldsymbol{\Phi}_{\beta_{\sf DAC}} \mathrm{diag}{\left\{ \bh_{{\sf D},k}\bh_{{\sf D},k}^{\sf H} \right\}} \bw_{i}.
    \label{eq:DAC_quantization_error_term}
\end{align}
Applying \eqref{eq:DAC_quantization_error_term} to \eqref{eq:R_Dk} with simplification, we have the reformulated SE of the $k$th DL user in \eqref{eq:R_Dk_refor} at the top of next page.
\begin{figure*}[t]
    \centering
    \begin{align}
    R_{{\sf D},k}^{\sf ref}({\bf W}) = \mathrm{log}_2 \left(1 + \frac{|\bh_{{\sf D},k}^{\sf H} \boldsymbol{\Phi}_{\alpha_{\sf DAC}} \bw_{k}|^{2}}{\sum_{i=1, i \ne k}^{K_{\sf D}} |\bh_{{\sf D},k}^{\sf H} \boldsymbol{\Phi}_{\alpha_{\sf DAC}} \bw_{i}|^2 +\sum_{i=1}^{K_{\sf D}} \! \bw^{\sf H}_{i} \boldsymbol{\Phi}_{\alpha_{\sf DAC}} \boldsymbol{\Phi}_{\beta_{\sf DAC}} \mathrm{diag}{\left\{ \bh_{{\sf D},k}\bh_{{\sf D},k}^{\sf H} \right\}} \bw_{i} + \|\bg_{{\sf CCI},k}\|^{2}\frac{P_{\sf U}}{P_{\sf D}} + \frac{\sigma_{\sf D}^2}{P_{\sf D}}} \right).
    \label{eq:R_Dk_refor}
    \end{align}
    \rule{\textwidth}{0.5pt}
\end{figure*}
We then define a weighted precoding vector as
\begin{align}
    \bv_{k} = \boldsymbol{\Phi}_{\alpha_{\sf DAC}}^{1/2} \bw_{k},
    \label{eq:wTov}
\end{align}
where $\bv_{k} \in \bbC^{N_t}$ for $k \in \{1,...,K_{\sf D}\}$. 
Stacking $\bv_{k}$, we obtain
\begin{align}
    \Bar{\bv} = \left[\bv_1^{\sf{T}}, \bv_2^{\sf{T}}, ..., \bv_{K_{\sf D}}^{\sf{T}}\right]^{\sf T} \in \bbC^{N_t K_{\sf D}}.\label{eq:stacked_v}
\end{align}
We assume $\|\Bar{\bv}\|^2 = 1$ which implies using maximum transmit power $P_{\sf D}$.
Based on $\|\bar\bv\|^2 = 1$, we rewrite the CCI and noise power in \eqref{eq:R_Dk_refor} that are the precoder-independent term as $\|\bg_{{\sf CCI},k}\|^{2}\frac{P_{\sf U}}{P_{\sf D}} + \frac{\sigma_{\sf D}^2}{P_{\sf D}}= \bar\bv^{\sf H} (\|\bg_{{\sf CCI},k}\|^{2}\frac{P_{\sf U}}{P_{\sf D}} + \frac{\sigma_{\sf D}^2}{P_{\sf D}} )\bar\bv$.

Let $\bM_{k}$ be 
\begin{align}
    \bM_{k} \!= \!(\boldsymbol{\Phi}_{\alpha_{\sf DAC}}^{1/2})^{\sf H} \bh_{{\sf D},k} \bh_{{\sf D},k}^{\sf H} \boldsymbol{\Phi}_{\alpha_{\sf DAC}}^{1/2} \! + \! \boldsymbol{\Phi}_{\beta_{\sf DAC}} \mathrm{diag}{\left\{ \bh_{{\sf D},k}\bh_{{\sf D},k}^{\sf H} \right\}}.
    \label{eq:Mk}
\end{align}
Using \eqref{eq:wTov}, \eqref{eq:stacked_v}, and \eqref{eq:Mk}, we represent $R^{\sf ref}_{{\sf D},k}(\bW)$ in \eqref{eq:R_Dk_refor} as a Rayleigh quotient form in logarithm:
\begin{align}
    \gamma_{{\sf D},k}(\bar {\bf v})  = \mathrm{log}_2 \left(\frac{\Bar{\bv}^{\sf H} \bA_{k} \Bar{\bv}}{\Bar{\bv}^{\sf H} \bB_{k} \Bar{\bv}} \right),\label{eq:Rayleigh_gamma_Dk}
\end{align}
where $\bA_{k} \in \bbC^{N_t K_{\sf D} \times N_t K_{\sf D}}$ and $\bB_{k} \in \bbC^{N_t K_{\sf D} \times N_t K_{\sf D}}$ are
\begin{align}
    &\bA_{k} = \; \mathbf{I}_{K_{\sf D}} \otimes \bM_{k} + \left( \|\bg_{{\sf CCI},k}\|^{2}\frac{P_{\sf U}}{P_{\sf D}} + \! \frac{\sigma_{\sf D}^2}{P_{\sf D}} \right)\bI_{N_t K_{\sf D}}, \label{eq:Ak}\\
    &\bB_{k} = \bA_{k} - \be_{k}\be_{k}^{\sf H} \otimes (\boldsymbol{\Phi}_{\alpha_{\sf DAC}}^{1/2})^{\sf H} \bh_{{\sf D},k} \bh_{{\sf D},k}^{\sf H} \boldsymbol{\Phi}_{\alpha_{\sf DAC}}^{1/2}. \label{eq:Bk} 
\end{align}

Next, we re-express $I_{{\sf SI},k}$ in \eqref{eq:I_SI} to facilitate the reformulation of the UL SE as
\begin{align}
    &\bar{I}_{{\sf SI},k} =\frac{I_{{\sf SI},k}}{P_{\sf U}} = \kappa_{\sf d} \frac{P_{\sf D}}{P_{\sf U}} \sum_{i=1}^{K_{\sf D}} \bw_{i}^{\sf H} \Psi_{{\sf SI},k} \bw_{i}, \label{eq:I_SI_tilde}
\end{align}
where 
\begin{align}
    \Psi_{{\sf SI},k} =& \, \boldsymbol{\Phi}_{\alpha_{\sf DAC}}^{\sf H} \bG_{\sf SI} \boldsymbol{\Phi}_{\alpha_{\sf ADC}}^{\sf H} \bff_{k} \bff_{k}^{\sf H} \boldsymbol{\Phi}_{\alpha_{\sf ADC}} \bG^{\sf H}_{\sf SI} \boldsymbol{\Phi}_{\alpha_{\sf DAC}} \nonumber\\
   &+ \, \boldsymbol{\Phi}_{\alpha_{\sf DAC}} \boldsymbol{\Phi}_{\beta_{\sf DAC}} \mathrm{diag}{\left\{\bG_{\sf SI} \boldsymbol{\Phi}_{\alpha_{\sf ADC}}^{\sf H}\bff_{k} \bff_{k}^{\sf H}\boldsymbol{\Phi}_{\alpha_{\sf ADC}} \bG^{\sf H}_{\sf SI}\right\}}.
   \label{eq:Psi_si}
\end{align}
We also rewrite the ADC quantization-related term in \eqref{eq:QN_ADC} as
\begin{align}
    \Bar{\mathrm{QN}}_{{\sf ADC},k} &= \frac{\mathrm{QN}_{{\sf ADC},k}}{P_{\sf U}} = \! \sum_{i=1}^{K_{\sf U}} \bh_{{\sf U},i}^{\sf H} \boldsymbol{\Phi}_{\alpha_{\sf ADC}} \boldsymbol{\Phi}_{\beta_{\sf ADC}} \mathrm{diag} \left\{\bff_{k}\bff_{k}^{\sf H} \right\} \bh_{{\sf U},i} \nonumber\\ 
    &+\frac{P_{\sf D}}{P_{\sf U}} \sum_{i=1}^{K_{\sf D}} \bw_{i}^{\sf H}  \Psi_{{\sf QN},k}\bw_{i}  \!+\! \frac{\sigma_{\sf U}^{2}}{P_{\sf U}} \bff_{k}^{\sf H} \boldsymbol{\Phi}_{\alpha_{\sf ADC}} \boldsymbol{\Phi}_{\beta_{\sf ADC}}\bff_{k}, \label{eq:ADC_quantization_error_term}
\end{align}
where
\begin{align}
    &\Psi_{{\sf QN},k} = \; \boldsymbol{\Phi}_{\alpha_{\sf DAC}}^{\sf H} \bG_{\sf SI} \boldsymbol{\Phi}_{\alpha_{\sf ADC}} \boldsymbol{\Phi}_{\beta_{\sf ADC}} \mathrm{diag}{\left\{ \bff_{k}\bff_{k}^{\sf H} \right\}} \bG^{\sf H}_{\sf SI} \boldsymbol{\Phi}_{\alpha_{\sf DAC}} \nonumber\\
    &\ + \boldsymbol{\Phi}_{\alpha_{\sf DAC}}\boldsymbol{\Phi}_{\beta_{\sf DAC}}\mathrm{diag}{\left\{\bG_{\sf SI} \boldsymbol{\Phi}_{\alpha_{\sf ADC}} \boldsymbol{\Phi}_{\beta_{\sf ADC}}\mathrm{diag}{\left\{\bff_{k}\bff_{k}^{\sf H}\right\}} \bG^{\sf H}_{\sf SI}\right\}}.
    \label{eq:Psi_qn}
\end{align}
We note that SI is included in the attenuated SI term $\Psi_{{\sf SI},k}$ in \eqref{eq:Psi_si} and in the ADC quantization noise-related term $ \Psi_{{\sf QN},k}$ in \eqref{eq:Psi_qn}.
Thus, the precoder needs to be designed to mitigate $\Psi_{{\sf SI},k}$ and $\Psi_{{\sf QN},k}$.
Since the digital SIC is not applied to  $\Psi_{{\sf QN},k}$, it is more critical in low-resolution ADC systems to reduce $\Psi_{{\sf QN},k}$.
Applying \eqref{eq:I_SI_tilde} and \eqref{eq:ADC_quantization_error_term} to \eqref{eq:R_Uk}, we also reorganize the UL SE in \eqref{eq:R_Uk_refor} at the top of the next page.
\begin{figure*}[t]
    \vspace{-1em}
    \centering
    \begin{align}
    R_{{\sf U},k}^{\sf ref}({\bf W},{\bf F}) = \mathrm{log_2}{\left( 1 + \frac{|\bff_{k}^{\sf H} \boldsymbol{\Phi}_{\alpha_{\sf ADC}} \bh_{{\sf U},k}|^{2}}{\sum_{i=1, i \ne k}^{K_{\sf U}} |\bff_k^{\sf H} \boldsymbol{\Phi}_{\alpha_{\sf ADC}} \bh_{{\sf U},i}|^{2} + \Bar{\mathrm{QN}}_{{\sf ADC},k} + \Bar{I}_{{\sf SI},k} + \|\bff_{k}^{\sf H} \boldsymbol{\Phi}_{\alpha_{\sf ADC}}\|^2\frac{\sigma_{\sf U}^2}{P_{\sf U}}}\right)}.
    \label{eq:R_Uk_refor}
    \end{align}
    \rule{\textwidth}{0.5pt}
\end{figure*}

Let $\bN_{k}$ be
\begin{align}
    \bN_{k} =& \; \frac{P_{\sf D}}{P_{\sf U}} \boldsymbol{(\Phi}_{\alpha_{\sf DAC}}^{-1/2})^{\sf H}\bigg(  \Psi_{{\sf QN},k} + \kappa_{\sf d} \Psi_{{\sf SI},k} \bigg)\boldsymbol{\Phi}_{\alpha_{\sf DAC}}^{-1/2}. \label{eq:Nk}   
\end{align}
Using \eqref{eq:wTov}, \eqref{eq:stacked_v} with $\|\Bar{\bv}\|^2 \! = \! 1$, 
we also rewrite $R^{\sf ref}_{{\sf U},k}(\bW, \bF)$ in \eqref{eq:R_Uk_refor} as a Rayleigh quotient form in logarithm:
\begin{align}
    \gamma_{{\sf U},k}(\bar {\bf v};{\bf F}) = \mathrm{log}_2 \left(\frac{\Bar{\bv}^{\sf H} \bC_{k} \Bar{\bv}}{\Bar{\bv}^{\sf H} \bD_{k} \Bar{\bv}} \right), \label{eq:Rayleigh_gamma_Uk}
\end{align}
where $\bC_{k} \in \bbC^{N_t K_{\sf D} \times N_t K_{\sf D}}$ and $\bD_{k} \in \bbC^{N_t K_{\sf D} \times N_t K_{\sf D}}$ are
\begin{align}
    &\bC_{k} = \mathbf{I}_{K_{\sf D}} \otimes \bN_k + \phi_{k} \bI_{N_t K_{\sf D}}, \label{eq:Ck}\\
    &\bD_{k} = \bC_{k} - \be_{k}\be_{k}^{\sf H} \otimes |\bff_{k}^{\sf H} \boldsymbol{\Phi}_{\alpha_{\sf ADC}} \bh_{{\sf U},k}|^{2}. \label{eq:Dk}
\end{align}
Here a precoder-independent term of \eqref{eq:R_Uk_refor} is represented as $\phi_{k} \!\! = \!\! \sum_{i=1}^{K_{\sf U}} \! |\bff_{k}^{\sf H} \boldsymbol{\Phi}_{\alpha_{\sf ADC}} \bh_{{\sf U},i}|^{2} \! + \! \sum_{i=1}^{K_{\sf U}} \! \bh_{{\sf U},i}^{\sf H} \! \boldsymbol{\Phi}_{\alpha_{\sf ADC}} \boldsymbol{\Phi}_{\beta_{\sf ADC}} \mathrm{diag} \{\bff_{k}\bff_{k}^{\sf H} \}  \bh_{{\sf U},i} \! + \! \bff_{k}^{\sf H} \boldsymbol{\Phi}_{\alpha_{\sf ADC}} \bff_{k} \frac{\sigma_{\sf U}^{2}}{P_{\sf U}}$.

Based on \eqref{eq:Rayleigh_gamma_Dk} and \eqref{eq:Rayleigh_gamma_Uk}, the optimization problem in \eqref{eq:objective} is reformulated with fixed ${\bf F}$ as
\begin{align}
    \label{eq:Rayleigh_objective}
    \underset{\Bar{\bv}}{\mathrm{maximize}}& \;\; R_{\Sigma}(\bar {\bf v}) = \sum_{k=1}^{K_{\sf D}} \gamma_{{\sf D},k}(\bar {\bf v})  + \sum_{k=1}^{K_{\sf U}} \gamma_{{\sf U},k} (\bar {\bf v};{\bf F}) 
    \\\label{eq:Rayleigh_constraint}
    \mathrm{subject \; to}& \;\; \|\Bar{\bv}\|^2 = 1.
\end{align}

%%%%%%%%%%%
\subsubsection{Precoding Algorithm}
\label{subsubsec:precoder}
We note that the non-convexity still remains in the reformulated problem.
Accordingly, we derive the first-order optimality condition of \eqref{eq:Rayleigh_objective} to identify stationary points of the problem.
\begin{lemma}
    \label{lem:KKT}
     The first-order optimality condition of  \eqref{eq:Rayleigh_objective} is represented in the generalized eigenvalue problem as
    \begin{align}
        \bB^{-1}_{\sf KKT}(\Bar{\bv}) \bA_{\sf KKT}(\Bar{\bv})\Bar{\bv} = \lambda(\Bar{\bv})\Bar{\bv}, 
        \label{eq:first_optimality_condition}
    \end{align}
    where
    \begin{align}
    \bA_{\sf KKT}(\Bar{\bv}) =& \left[ \sum_{k=1}^{\sf K_{\sf D}} \left( \frac{\bA_{k}}{\Bar{\bv}^{\sf H} \bA_{k} \Bar{\bv}}\right) + \sum_{k=1}^{\sf K_{\sf U}} \left( \frac{\bC_{k}}{\Bar{\bv}^{\sf H} \bC_{k} \Bar{\bv}}\right) \right] \lambda_{\sf num}(\Bar{\bv}), \label{eq:A_KKT}\\
    \bB_{\sf KKT}(\Bar{\bv}) =& \left[ \sum_{k=1}^{\sf K_{\sf D}} \left( \frac{\bB_{k}}{\Bar{\bv}^{\sf H} \bB_{k} \Bar{\bv}}\right) + \sum_{k=1}^{\sf K_{\sf U}} \left( \frac{\bD_{k}}{\Bar{\bv}^{\sf H} \bD_{k} \Bar{\bv}}\right) \right] \lambda_{\sf den}(\Bar{\bv}),\label{eq:B_KKT}\\
    \lambda(\Bar{\bv}) =&\; \prod_{k=1}^{K_{\sf D}} \left( \frac{\Bar{\bv}^{\sf H}\bA_{k}\Bar{\bv}}{\Bar{\bv}^{\sf H} \bB_{k} \Bar{\bv}}\right) \prod_{k=1}^{K_{\sf U}} \left( \frac{\Bar{\bv}^{\sf H}\bC_{k}\Bar{\bv}}{\Bar{\bv}^{\sf H} \bD_{k} \Bar{\bv}}\right),\label{eq:lambda_v}
\end{align}
and $\lambda_{\sf num}(\Bar{\bv})$ and $\lambda_{\sf den}(\Bar{\bv})$  can be any functions that satisfy $\lambda(\Bar{\bv})=\lambda_{\sf num}(\Bar{\bv})/\lambda_{\sf den}(\Bar{\bv})$.

\begin{proof}
See Appendix~\ref{app:KKT}.
\end{proof}
\end{lemma}

%%%%% GPI 알고리즘 설명 %%%%%
Note that the eigenvalue $\lambda(\bar \bv)$ relates to the objective function in \eqref{eq:Rayleigh_objective} as $R_{\Sigma}(\bar {\bv}) \! = \! \log_2 \lambda (\bar \bv)$.
According to Lemma~\ref{lem:KKT}, the stationary points satisfy \eqref{eq:first_optimality_condition} as an eigenvector of $\bB^{-1}_{\sf KKT}(\Bar{\bv}) \bA_{\sf KKT}(\Bar{\bv})$, and the corresponding eigenvalue is $\lambda(\bar \bv)$ which is proportional to the sum SE $R_{\Sigma}(\bar \bv)$.
This interpretation implies that we need to find the principal eigenvector of the problem \eqref{eq:first_optimality_condition} for maximizing $R_{\Sigma}(\bar \bv)$, and the principal eigenvector is the precoding vector on one of the stationary points, which is equivalent to finding the best local optimal solution of \eqref{eq:objective}. 
Accordingly, the first-order optimality condition \eqref{eq:first_optimality_condition} is equivalent to the NEPv, and we leverage the GPI method to find its principal eigenvector.

We present the sum SE maximization precoding method in Algorithm~\ref{alg:one}.
The key steps of Algorithm~\ref{alg:one} are as follows:
we first initialize the stacked precoding vectors $\Bar{\bv}^{(0)}$ using a conventional linear precoder such as MRT, ZF, and regularized ZF (RZF).
At each iteration $n$, we build the matrices $\bA_{\sf KKT}(\Bar{\bv}^{(n-1)})$ and $\bB_{\sf KKT}(\Bar{\bv}^{(n-1)})$ according to $\eqref{eq:A_KKT}$ and $\eqref{eq:B_KKT}$.
Then, we compute $\Bar{\bv}^{(n)}$ as $\Bar{\bv}^{(n)} \! = \! \bB_{\sf KKT}^{-1}(\Bar{\bv}^{(n-1)}) \bA_{\sf KKT}(\Bar{\bv}^{(n-1)}) \Bar{\bv}^{(n-1)}$ followed by normalization $\Bar{\bv}^{(n)} \! = \! \Bar{\bv}^{(n)} \! / \|\Bar{\bv}^{(n)}\|$. 
This normalization ensures the transmit power constraint.
We repeat this process until the stopping criterion $\|\Bar{\bv}^{(n)} \! - \! \Bar{\bv}^{(n-1)}\| < \varepsilon_{\sf \bv}$ is met, where $\varepsilon_{\sf \bv} > 0$ denotes a tolerance level or until the iteration count reaches the maximum count $n > n_{\sf max}$.

%%%%% Precoder Algorithm %%%%%
\begin{algorithm}[!t]
\caption{Precoder Design}\label{alg:one}
\bf{initialize}: $\Bar{\bv}^{(0)}$\\
\rm{Set the iteration count} $n=1.$\\
\While{$\|\Bar{\bv}^{(n)} - \Bar{\bv}^{(n-1)}\| > \varepsilon_{\bv}$ \rm{and} $n \leq n_{\sf max}$}{
\rm{Build matrix} $\bA_{\sf KKT}(\Bar{\bv}^{(n-1)})$ in \eqref{eq:A_KKT}.\\
\rm{Build matrix} $\bB_{\sf KKT}(\Bar{\bv}^{(n-1)})$ in \eqref{eq:B_KKT}.\\
\rm{Compute} $\Bar{\bv}^{(n)}$ = $\bB_{\sf KKT}^{-1}(\Bar{\bv}^{(n-1)}) \bA_{\sf KKT}(\Bar{\bv}^{(n-1)}) \Bar{\bv}^{(n-1)}.$\\
Normalize $\Bar{\bv}^{(n)} = \Bar{\bv}^{(n)}/\|\Bar{\bv}^{(n)}\|.$\\
$n \gets n+1.$\\
}
$\Bar{\bv}^{\star} \gets \Bar{\bv}^{(n)}.$\\
\Return $\Bar{\bv}^{\star}$
\end{algorithm}
%%%%%%%%%%%%%%%%%%%%%%%%%%%%%%%

% %%%%% UL에서의 Q-MMSE realization %%%%%
% \subsection{Combiner Optimization}
% \label{subsubsec:combiner}
Regarding the combiner optimization, it is known that the MMSE equalizer maximizes the received SINR. 
Accordingly, we build a quantization-aware linear MMSE (qMMSE) equalizer with the derived precoder from Algorithm~\ref{alg:one}.
From \eqref{eq:r_Uq}, we set the interference plus noise term for UL as
\begin{align}
    \bz_{k} =& \sum_{i=1, i \ne k}^{K_{\sf U}} \boldsymbol{\Phi}_{\alpha_{\sf ADC}} \bh_{{\sf U},i} s_{{\sf U},i} + \sqrt{\kappa_{\sf d}} \sum_{i=1}^{K_{\sf D}} \boldsymbol{\Phi}_{\alpha_{\sf ADC}} \bG^{\sf H}_{\sf SI} \boldsymbol{\Phi}_{\alpha_{\sf DAC}} \bw_{i} s_{{\sf D},i} \nonumber \\
    &+ \sqrt{\kappa_{\sf d}} \boldsymbol{\Phi}_{\alpha_{\sf ADC}} \bG^{\sf H}_{\sf SI} \bq_{\sf DAC} + \bq_{\sf ADC} + \boldsymbol{\Phi}_{\alpha_{\sf ADC}} \bn_{\sf U}. \label{eq:colored_noise}
\end{align}
Using \eqref{eq:colored_noise}, the combining vector is defined as
\begin{align}
    \label{eq:MMSE_receiver}
    {\bf f}_{k} =  \bK_{\bz_{k}}^{-1} \boldsymbol{\Phi}_{\alpha_{\sf ADC}} \bh_{{\sf U},k},
\end{align}
where $\bK_{\bz_{k}} = \bbE \left[\bz_{k} \bz_{k}^{\sf H} \right]$.

%%%%% Q-GPI-FD Algorithm %%%%%
\subsection{Joint FD Beamforming Algorithm}
\begin{algorithm}[!t]
\caption{Proposed Algorithm}\label{alg:two}
% \caption{Joint Beamforming Method for Quantized FD MU-MISO Systems}\label{alg:two}
\bf{initialize}: $\Bar{\bu}^{(0)}, \Bar{\bv}^{(0)}, \Bar{\bff}^{(0)}$\\
\rm{Set the iteration count} $t=1.$\\
\While{$\|\Bar{\bu}^{(t)} - \Bar{\bu}^{(t-1)}\| > \varepsilon_{\bu}$ \rm{or}  $\|\Bar{\bff}^{(t)} - \Bar{\bff}^{(t-1)}\| / \|\Bar{\bff}^{(t)}\|> \varepsilon_{\bff}$ \rm{and} $t \leq t_{\sf max}$}{
Find $\Bar{\bv} \gets$ Algorithm~\ref{alg:one}($\bar\bff^{(t-1)};\bH_{\sf D}, \bG_{\sf CCI}$).\\
\rm{Compute qMMSE} $\Bar{\bff}^{(t)}$ in \eqref{eq:MMSE_receiver} by using  \eqref{eq:wTov}.\\
Set $\bar{\bu}^{(t)} \gets \bar{\bv}$.\\
Set $t \gets t+1.$
}
$\Bar{\bv}^{\star} \gets \Bar{\bu}^{(t)}$ and $\bff^{\star} \gets \Bar{\bff}^{(t)}$.\\
\Return $\bV^{\star} \! = \! \left[\bv_{1}^{\star}, \bv_{2}^{\star}, ... ,\bv_{K_{\sf D}}^{\star} \right]$ and $\bF^{\star} \! = \! \left[\bff_{1}^{\star}, \bff_{2}^{\star}, ... ,\bff_{K_{\sf U}}^{\star} \right]$
\end{algorithm}
%%%%%%%%%%%%%%%%%%%%%%%%%%%%%%

The joint FD beamformer is alternately optimized by using Algorithm~\ref{alg:one} and the qMMSE combiner in \eqref{eq:MMSE_receiver}.
The steps of the proposed algorithm are as follows:
we first initialize the stacked precoding vectors $\Bar{\bv}^{(0)}$ and $\Bar{\bu}^{(0)}$ and combining vector $\Bar{\bff}^{(0)}$ using conventional linear beamformers, 
% and the combining vector $\Bar{\bff}^{(0)}$,
where we define $\Bar{\bff} = [\bff_{1}^{\sf T}, ..., \bff_{K_{\sf U}}^{\sf T}]^{\sf T}$ for notational consistency with the precoding vector. 
At each iteration $t$, we derive $\Bar{\bv}^{(n)}$ from Algorithm~\ref{alg:one}.
Then, we build ${\bw}_k$ by using \eqref{eq:wTov} with $\Bar{\bv}^{(n)}$, and compute $\Bar{\bff}^{(t)}$.
We repeat this process until $\|\Bar{\bu}^{(t)} \! - \! \Bar{\bu}^{(t-1)}\| \! < \! \varepsilon_{\sf \bu}$ and $\|\Bar{\bff}^{(t)} \! - \Bar{\bff}^{(t-1)}\|/\|\Bar{\bff}^{(t)}\| < \varepsilon_{\sf \bff}$, where $\varepsilon_{\sf \bu} > 0$ and $\varepsilon_{\sf \bff} > 0$ denote a tolerance level or until $t > t_{\sf max}$.

The computational complexity of the proposed algorithm is primarily determined by the matrix inversion $\bB_{\sf KKT}^{-1}(\Bar{\bv})$ in Algorithm~\ref{alg:one}.
Although $\bB_{\sf KKT}(\Bar{\bv})$ is an $N_tK_{\sf D} \times N_tK_{\sf D}$ matrix, thanks to its block-diagonal structure, we only need $\CMcal{O} \left(N_{t}^{3} K_{\sf D} \right)$ instead of $\CMcal{O} \left((N_{t} K_{\sf D})^3 \right)$ by computing the inverses of $K_{\sf D}$ sub-matrices in $\bB_{\sf KKT}(\Bar{\bv})$ separately \cite{choi:twc:19}.
In this regard, the total complexity is $\CMcal{O} \left(T N_{t}^{3} K_{\sf D} \right)$, where $T$ is the total number of iterations of the power method in Algorithm~\ref{alg:one} including the outer loop alternation.
We remark that the complexity of the proposed algorithm is lower than that of the state-of-the-art minorization-maximization (MM) algorithm in \cite{kim:tvt:16}. 
The MM algorithm has a complexity of $\CMcal{O} \left( T_{\sf MM} (N_r K_{\sf U})^{4.5} \right)$ for UL and $\CMcal{O} \left( T_{\sf MM} (N_t K_{\sf D})^{4.5} \right)$ for DL. 
Accordingly, the total computational complexity is $\CMcal{O} \left( T_{\sf MM} ((N_r K_{\sf U})^{4.5} \! + \! (N_t K_{\sf D})^{4.5}) \right)$, where $T_{\sf MM}$ is total iterations of the MM algorithm in \cite{kim:tvt:16}.

%%%%% Numerical Results %%%%%
\section{Numerical Analysis}
\label{sec:numerical}
\begin{figure}[t]
    \centering
    \includegraphics[width=0.65\linewidth]{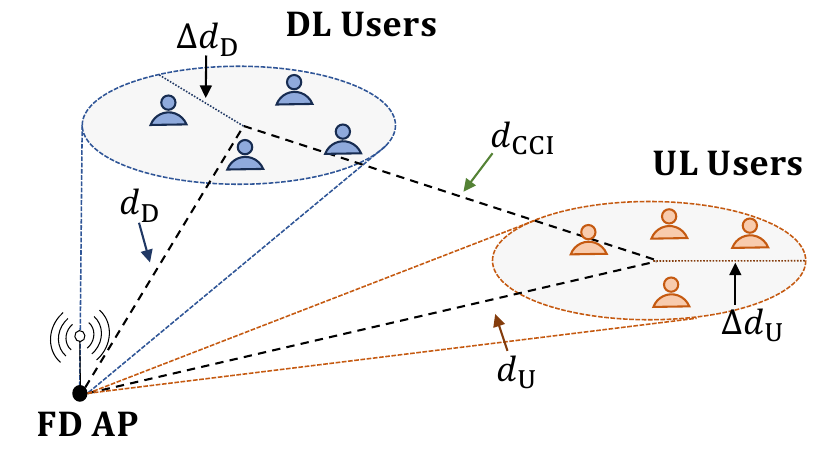}
    \caption{The considered FD MU-MISO wireless networks in simulations.}
    \label{fig:Scenario}
        \vspace{-1em}
\end{figure}

In this section, we validate the derived analytical results in Section~\ref{sec:ADC} and evaluate the performance of the proposed method in Section~\ref{sec:beamforming} by comparing with benchmarks.
As shown in Fig.~\ref{fig:Scenario}, the DL and UL users are randomly distributed within circles of radius $\Delta d_{\sf D} \! = \! \Delta d_{\sf U} \! = \! 4\,\mathrm{m}$, centered at distances $d_{\sf D}\! = \! d_{\sf U} \! = \! 15\,\mathrm{m}$ from the AP.
Unless otherwise specified, the inter-group distance is fixed at $d_{\sf CCI} \! = \! 30\,\mathrm{m}$.
We adopt the one-ring channel model \cite{adhikary:tit:13} to construct the small-scale channel covariances $\bC_{{\sf D},k}$ and $\bC_{{\sf U},k}$, respectively.
The close-in path-loss model \cite{sun:vtc83:2016} is employed for DL, UL, and CCI channels under an open square scenario, with a uniform path-loss value $\rho_{{\sf CCI},k} \! = \! \rho_{\sf CCI}$ assumed for simplicity.
Our setup involves $10\,\mathrm{GHz}$ carrier frequency, $500\,\mathrm{MHz}$ bandwidth, $5\,\mathrm{dB}$ noise figure, $8.4\,\mathrm{dB}$ shadow fading standard deviation, $2.8$ path-loss exponent, and noise power spectral density of $-174\,\mathrm{dBm/Hz}$.
% ---------- figure: system model -----------
% -------------------------------------------
In the proposed algorithm, we set $\epsilon_{\bu} \! = \epsilon_{\bv} \! = \epsilon_{\bff} \! = \! 10^{-2}$ and $t_{\sf max} \! = \! n_{\sf max} \! = \! 30$.
The residual SI after analog SIC is assumed to be suppressed by digital SIC down to the UL noise floor, i.e., $P_{\sf D} \kappa_{\sf a}\kappa_{\sf d} = \sigma_{\sf U}^2$.
To construct HD baselines, we disable SIC capabilities and CCI power by setting $\kappa_{\sf a} \! = \! \kappa_{\sf d} \! = \! 0$ and $\rho_{\sf CCI} \! = \! 0$, respectively.
Then, we compute the sum of HD DL and UL SEs as  $\sum_{k=1}^{K_{\sf D}} R^{\sf HD}_{{\sf D},k} \! = \! (1-\lambda) \sum_{k=1}^{K_{\sf D}} R_{{\sf D},k}(\bW)$ and $\sum_{k=1}^{K_{\sf U}} R^{\sf HD}_{{\sf U},k} \! = \! \lambda \sum_{k=1}^{K_{\sf U}} R_{{\sf U},k}(\bW, \bF)$, where we assume $\lambda \! = \! 1/2$.
We now consider the following FD and HD benchmarks:
\begin{itemize}
    \item {\bf (FD) qMM-qMMSE:} The FD optimization algorithm derived by modifying the MM precoding in \cite{kim:tvt:16} to incorporate quantization errors and adopting qMMSE.
    \item {\bf (FD) qRZF-qMMSE:} The linear FD scheme employing quantization-aware RZF (qRZF) precoding and qMMSE combining.
    \item {\bf (HD) qGPI $\&$ qMMSE}: The HD scheme that uses a quantization-aware GPI (qGPI) based on the power method in \cite{choi:twc:19} for precoding and qMMSE for combining.
    \item {\bf (HD) qWMMSE $\&$ qMMSE}: The HD method that uses a quantization-aware weighted MMSE (qWMMSE) based approach \cite{christensen:twc:2008} for precoding and qMMSE for combining.
    \item {\bf (HD) qWMMSE-AO $\&$ qMMSE}: The HD scheme that uses a qWMMSE alternating optimization (qWMMSE-AO) approach inspired by successive convex approximation in \cite{razaviyayn:Dr:2014} for precoding and qMMSE for combining.
    \item {\bf (HD) qRZF $\&$ qMMSE}: The HD method that utilizes the qRZF for precoding and qMMSE for combining.
\end{itemize}
We remark that all benchmarks are modified to incorporate the DAC and ADC quantization effects based on the considered system for more fair comparison in terms of SE performance.

\subsection{SQNR}
\label{subsec:SQNRanalysis}
%%%%% Ergodic SQNR vs ADCbits %%%%%
\begin{figure}[t]
    \centering
    \subfigure[$K_{\sf D}=K_{\sf U}=1$]{
    \includegraphics[width=0.85\linewidth]{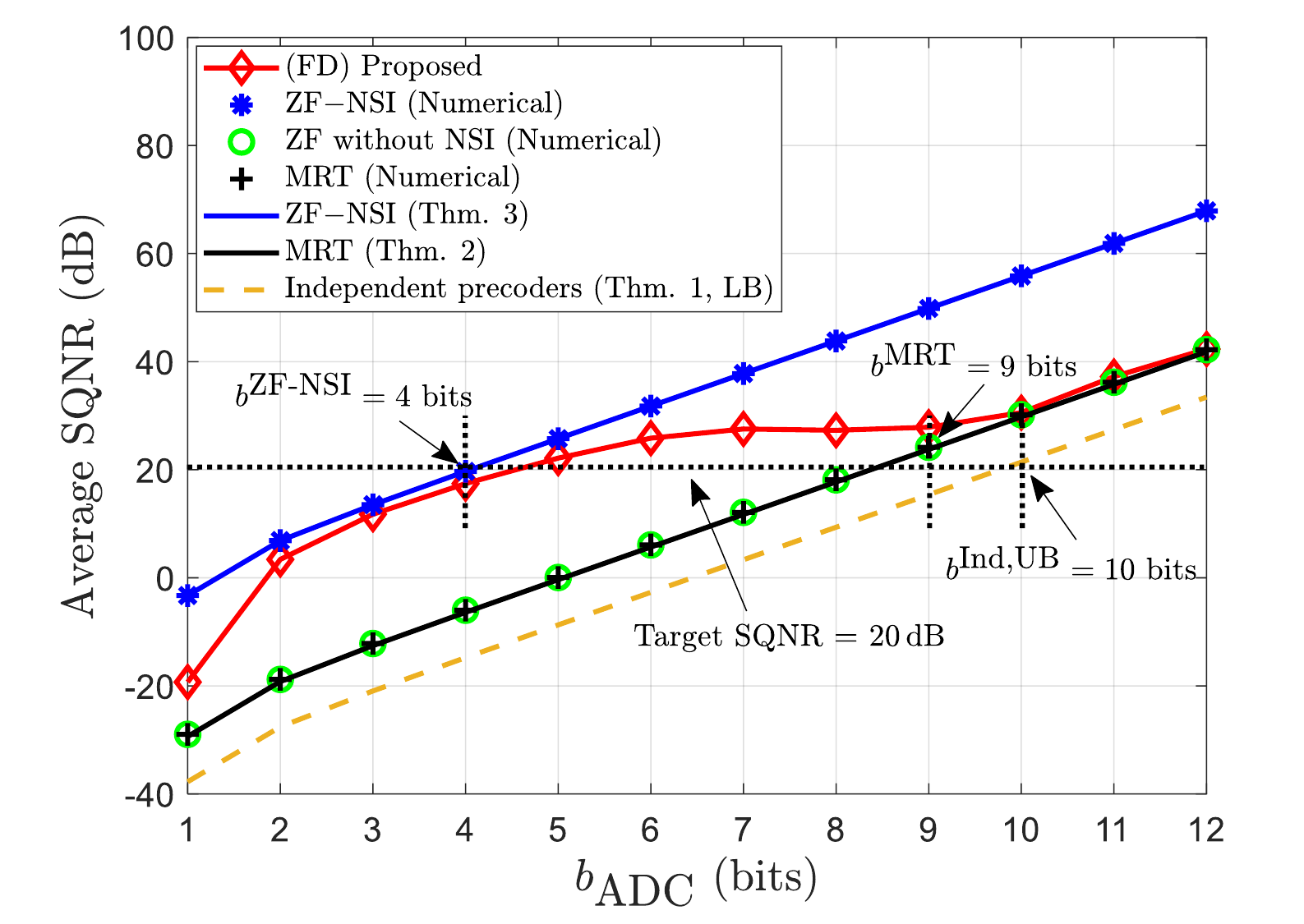}
    \label{fig:Avg_SQNRvsADCbits}
    }
    \subfigure[$K_{\sf D}=K_{\sf U}=2$]{
    \includegraphics[width=0.85\linewidth]{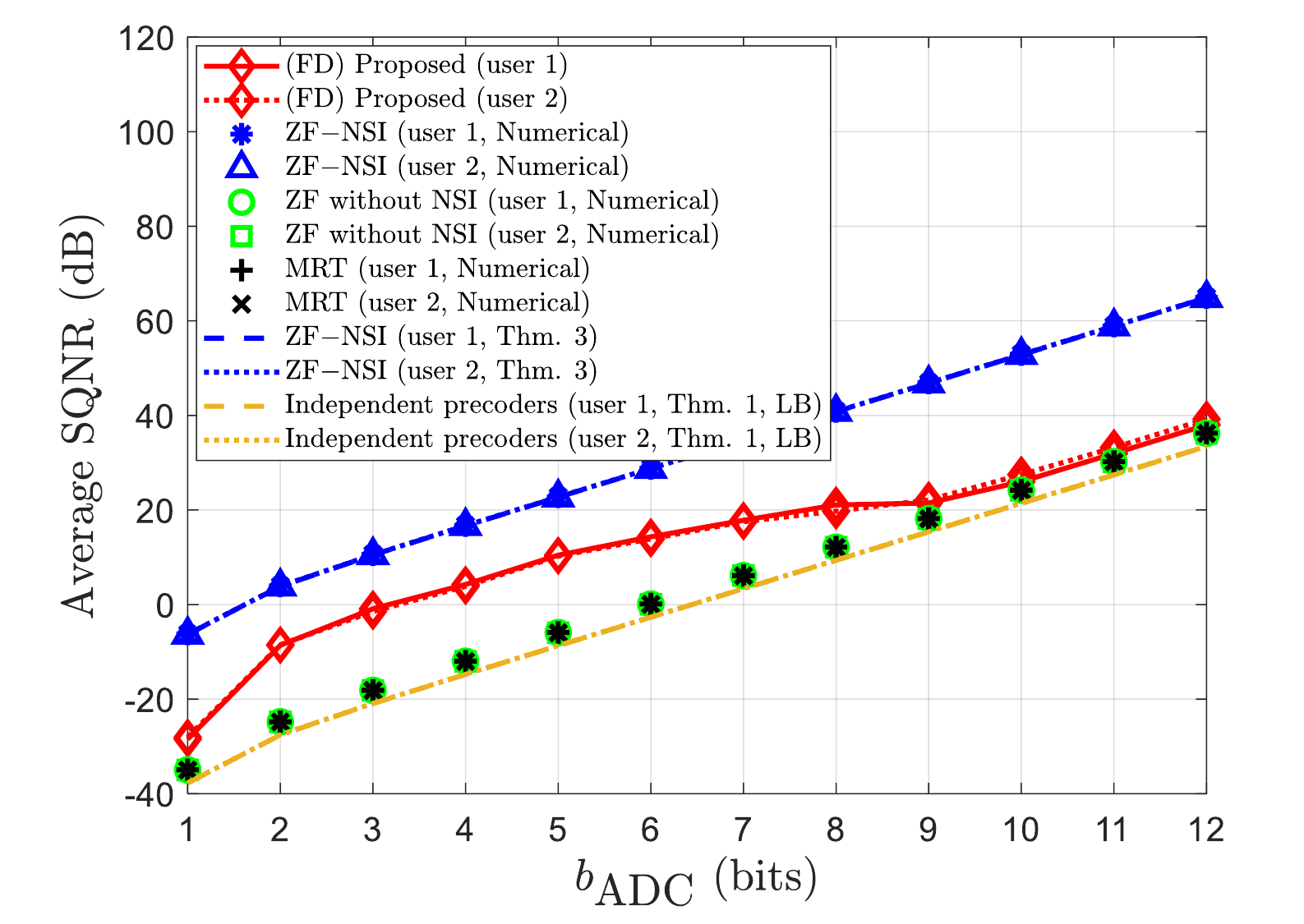}
    \label{fig:Avg_SQNRvsADCbits_multi}
    }
    \caption{{The UL average SQNR versus the number of ADC  bits  for $N_t=N_r=32$, $P_{\sf D}=24\,\mathrm{dBm}$, $P_{\sf U}=23\,\mathrm{dBm}$, $\kappa_{\sf a}=-60\,\mathrm{dB}$, and $b_{{\sf DAC},m}=\infty$ bits}}
    \label{fig:Avg_SQNR}
    \vspace{-1em}
\end{figure}
In Fig.~\ref{fig:Avg_SQNRvsADCbits}, we first validate the SQNR and bit analysis results.
We set $\kappa_{\sf a} \! = \! -60\,\mathrm{dB}$, which is typically regarded as a maximum analog SIC capability \cite{kim2024learning} in a small cell scenario.
Fig.~\ref{fig:Avg_SQNRvsADCbits} shows the UL average SQNR versus the number of ADC quantization bits for the case of $K_{\sf D} \! = \! K_{\sf U} \! = \! 1$.
In Fig.~\ref{fig:Avg_SQNRvsADCbits}, the numerical results for all linear schemes match their closed-form expressions derived in the theorems, for which Theorem~\ref{Th:theorem2} also provides a lower bound.
To achieve a UL average SQNR of $20\, \mathrm{dB}$, for example, the proposed method requires only about 5 ADC bits, compared to $b^{\sf ZF-NSI}= 4$ bits for ZF-NSI, approximately $b^{\sf MRT} = 9$ bits for MRT, and the theoretical upper bound of about $b^{\sf Ind, UB} =10$ bits for SI-independent precoders. 
Thus, we conjecture that the proposed precoder is able to balance the SI suppression and DL SE maximization as its corresponding ADC bit follows Proposition~\ref{prop:proposition1}.
To be specific, in the low-resolution regime, the proposed method achieves the SQNR comparable to that of ZF-NSI, while in the high-resolution regime its SQNR converges to that of MRT.
This indicates that the proposed scheme prioritizes mitigating residual SI (equivalently, quantization distortion) when fewer ADC bits are available.
However, as quantization errors become negligible with higher ADC bits, the design objective shifts toward maximizing DL SE.
Similar interpretation also holds for Fig.~\ref{fig:Avg_SQNRvsADCbits_multi} which shows the UL average SQNR versus the number of ADC  bits for the case of $K_{\sf D}=K_{\sf U}=2$.

\subsection{Spectral Efficiency}
\label{subsec:sim_SE}
Now, we compare the proposed method and other baselines in terms of the sum SE.
%%%%% SEvsADCbits %%%%%
\begin{figure}[!t]
    \centering
    \subfigure[]{
    \includegraphics[width=0.815\linewidth]{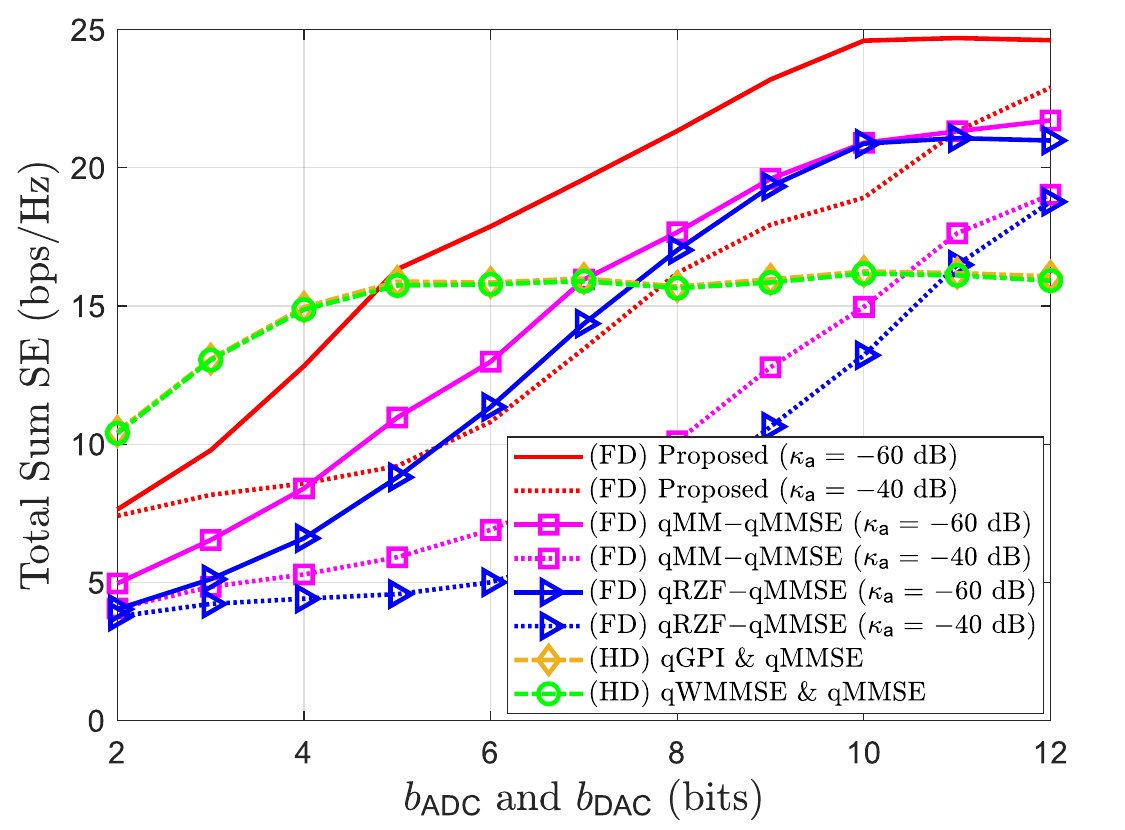}
    \label{fig:SEvsbits}
    }
    \subfigure[]{
    \includegraphics[width=0.815\linewidth]{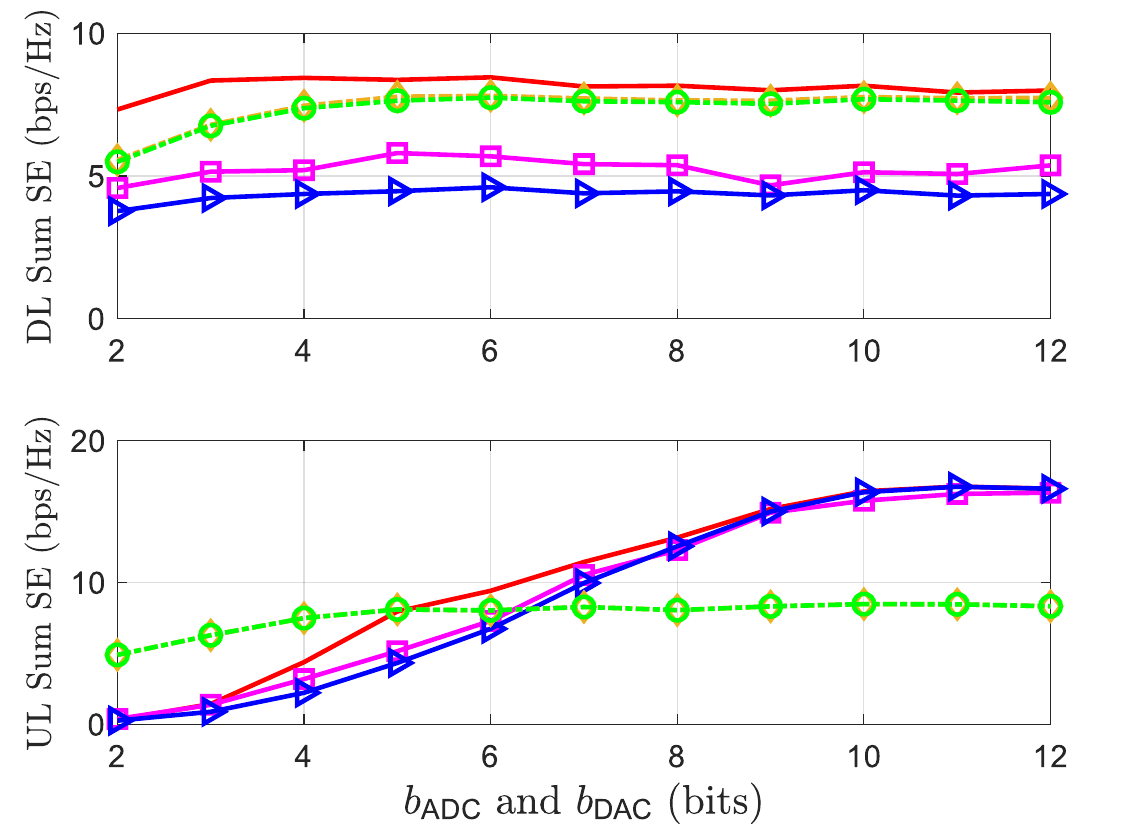}
    \label{fig:SEvsbits_DLnUL}
    }
    \caption{(a) The total sum SE with analog SIC capability $\kappa_{\sf a} \in \{-40, -60\}\,\mathrm{dB}$, (b) DL and UL sum SEs  with $\kappa_{\sf a} = -60\,\mathrm{dB}$ versus the number of ADC and DAC bits  for $N_t=N_r=16$, $K_{\sf D}=K_{\sf U}=4$, $P_{\sf D}=24\,\mathrm{dBm}$, $P_{\sf U}=23\,\mathrm{dBm}$.}
    \label{fig:SEvsADCbits}
        \vspace{-1em}
\end{figure}
%%%%%%%%%%%%%%%%%%%%%%%
Fig.~\ref{fig:SEvsbits} shows the total sum SE versus ADC and DAC quantization bits.
It is observed that the proposed method consistently outperforms the FD benchmarks.
For $\kappa_{\sf a} \! = \! -60\,\mathrm{dB}$, the proposed method shows  noticeable gains over the HD cases when $b_{\sf ADC} \! = \! b_{\sf DAC} \! \geq \! 6$ bits.
As noted in  Fig.~\ref{fig:SEvsbits_DLnUL}, this is mainly because the UL sum SE is highly degraded under $6$ bits due to the residual analog SI significantly contributing to the quantization noise.
This behavior differs from the conventional HD systems, where 4-bit ADCs are typically sufficient \cite{choi:twc:2022, studer2016quantized, verenzuela2016hardware, jacobsson2017throughput}.
These results highlight that enhancing the analog SIC capability is essential for improving sum SE as much as possible under coarse quantization.
We note that the FD baselines require more bits to outperform the HD systems, i.e., $b_{\sf ADC} \! = \! b_{\sf DAC} \! \geq \! 8$ bits with $\kappa_{\sf a} \! = \! -60\, \mathrm{dB}$, than the proposed method because of insufficient control over the residual analog SI, quantization error, and IUI.
This reveals that careful optimization of the beamforming can reduce the required number of ADC bits, thereby offering more opportunities for enhancing SE and EE.

\begin{figure}[t]
    \centering
    \includegraphics[width=0.815\linewidth]{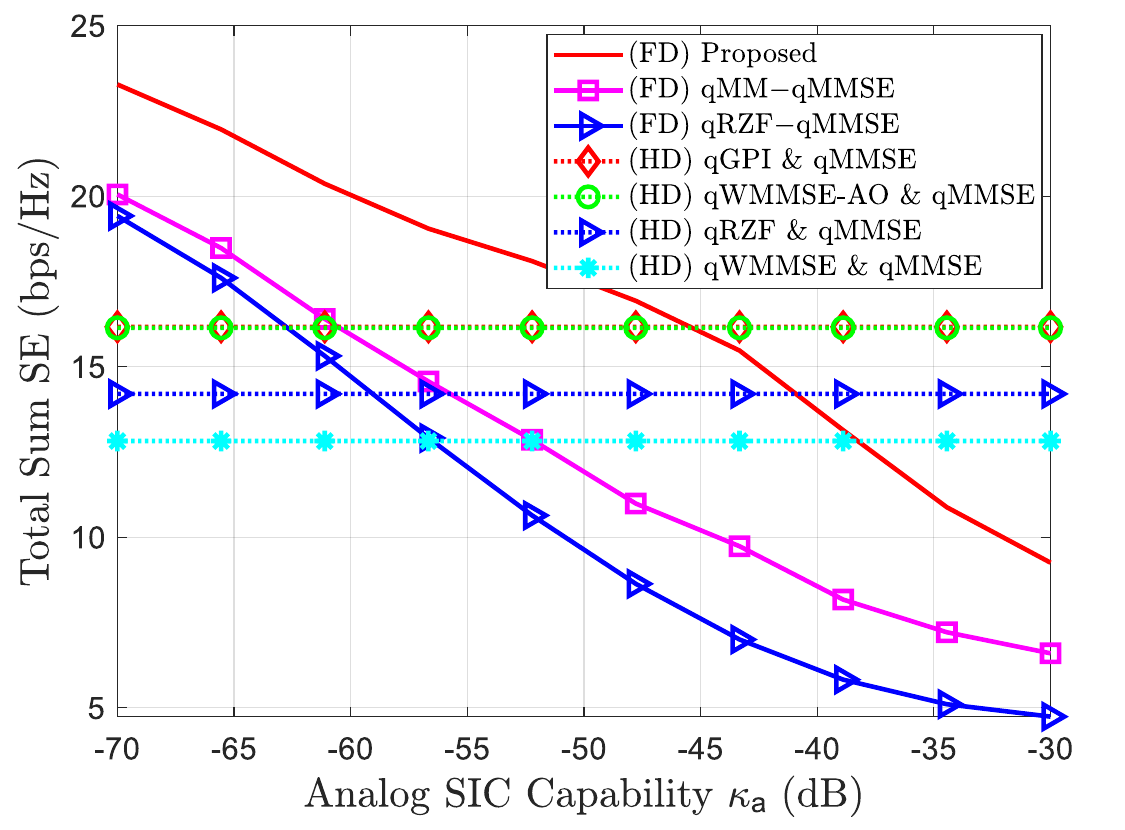}
    \caption{The total sum SE versus the analog SIC capability for $N_t=N_r=16$, $K_{\sf D}=K_{\sf U}=4$, $P_{\sf D}=24\,\mathrm{dBm}$, $P_{\sf U}=23\,\mathrm{dBm}$, and $b_{{\sf ADC},n}=b_{{\sf DAC},m}=7$ bits.}
    \label{fig:SIC_total}
        \vspace{-1em}
\end{figure}

Fig.~\ref{fig:SIC_total} illustrates the total sum SE versus the analog SIC capability.
As the analog SIC capability degrades, the total sum SE of all FD algorithms declines.
With 7-bit quantization, the proposed method achieves a noticeable SE gain compared to HD systems when the analog SIC capability is $\kappa_{\sf a} \! < \! -50\,\mathrm{dB}$. 
This indicates that the analog SIC capability has a significant impact on the FD systems with coarse ADC quantization as it directly determines the quantization noise power.
The same insight is observed in Fig.~\ref{fig:SEvsADCbits}, highlighting the need to improve the beamforming design and analog SIC capability to enable low-resolution quantizers.

%%%%% Analog SIC %%%%%%%%%%%%%%%
%%%%%%%%%%%%%%%%%%%%%%

%%%%%%%%%%%%%%% Antenna %%%%%%%%%%%%%%%
\begin{figure}[t]
    \centering
    \includegraphics[width=0.815\linewidth]{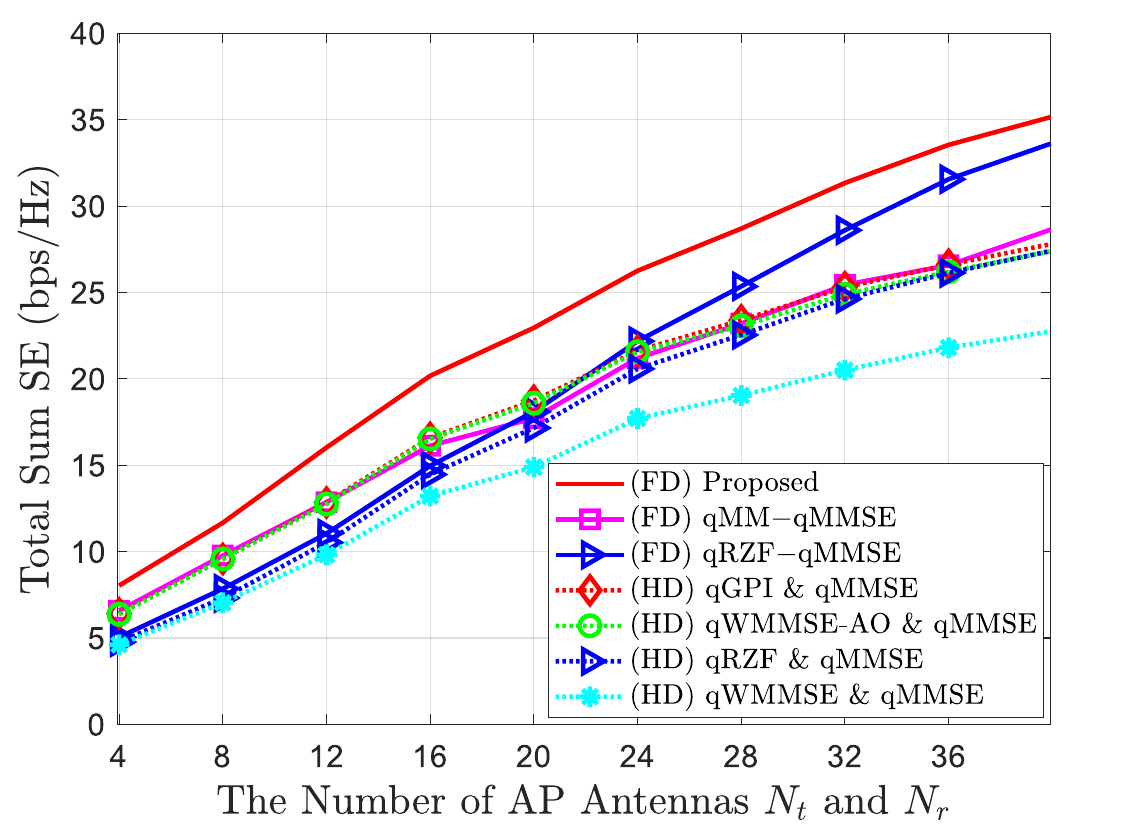}
    \caption{The total sum SE versus the number of AP antennas for $K_{\sf D}=K_{\sf U}=4$, $P_{\sf D}=24\,\mathrm{dBm}$, $P_{\sf U}=23\,\mathrm{dBm}$, $\kappa_{\sf a}=-60\,\mathrm{dB}$, and $b_{{\sf ADC},n}=b_{{\sf DAC},m}=7$ bits.}
    \label{fig:Antenna_total}
        \vspace{-1em}
\end{figure}
%%%%%%%%%%%%%%%%%%%%%%%%%%%%%%%%%%%%%%%
Fig.~\ref{fig:Antenna_total} shows the total sum SE versus the number of AP's transmit and receive antennas.
The proposed algorithm attains the highest sum SE among all FD and HD benchmarks.
As the number of antennas increases, the FD baselines exhibit higher improvement from the HD systems.
This suggests that the FD precoding with a larger number of antennas contributes to improving the sum SE by not only increasing the DL SINR but also decreasing the SI. 
Thus, it is more important to have larger antenna arrays with an optimized SI-aware beamformer in the quantized FD systems than the quantized HD systems.

%%%%% transmit power change %%%%%
\begin{figure}[t]
    \centering
    \includegraphics[width=0.815\linewidth]{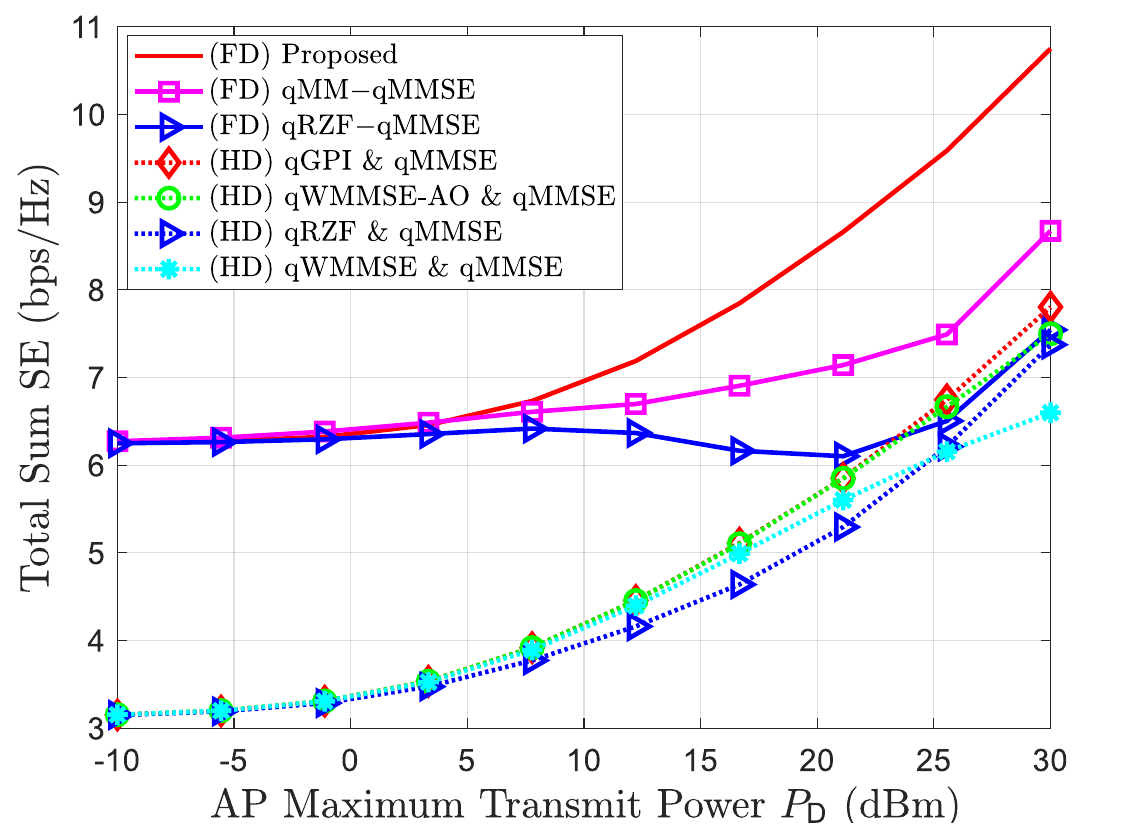}
    \caption{The total sum SE versus the AP transmit power  for $N_t=N_r=4$, $K_{\sf D}=K_{\sf U}=2$, $P_{\sf U}=23\,\mathrm{dBm}$, $\kappa_{\sf a} =-60\,\mathrm{dB}$, and $b_{{\sf ADC},n}=b_{{\sf DAC},m}=7$ bits.}
    \label{fig:P_D}
        % \vspace{-1em}
\end{figure}
%%%%%%%%%%%%%%%%%%%%%%%%%%%%%%%%%
Fig.~\ref{fig:P_D} shows the total sum SE versus the AP's maximum transmit power $P_{\sf D}$.
Over the entire range of $P_{\sf D}$, the proposed method exhibits the highest sum SE.
As $P_{\sf D}$ increases, the SE gain of the FD baselines over HD systems diminishes, because $P_{\sf D}$ leads to stronger residual analog SI after the analog SIC, thereby elevating the quantization error floor. 
In contrast, the proposed method maintains its SE advantage by jointly suppressing the residual SI while sustaining high DL SE.
These results underscore that, in quantized FD architectures, an advanced SI-aware precoder that maximizes DL SE is essential for effectively leveraging higher transmit power.

%%%%%%%%%%% DL_convergence %%%%%%%%%%%%
\begin{figure}[t]
    \centering
    \includegraphics[width=0.775\linewidth]{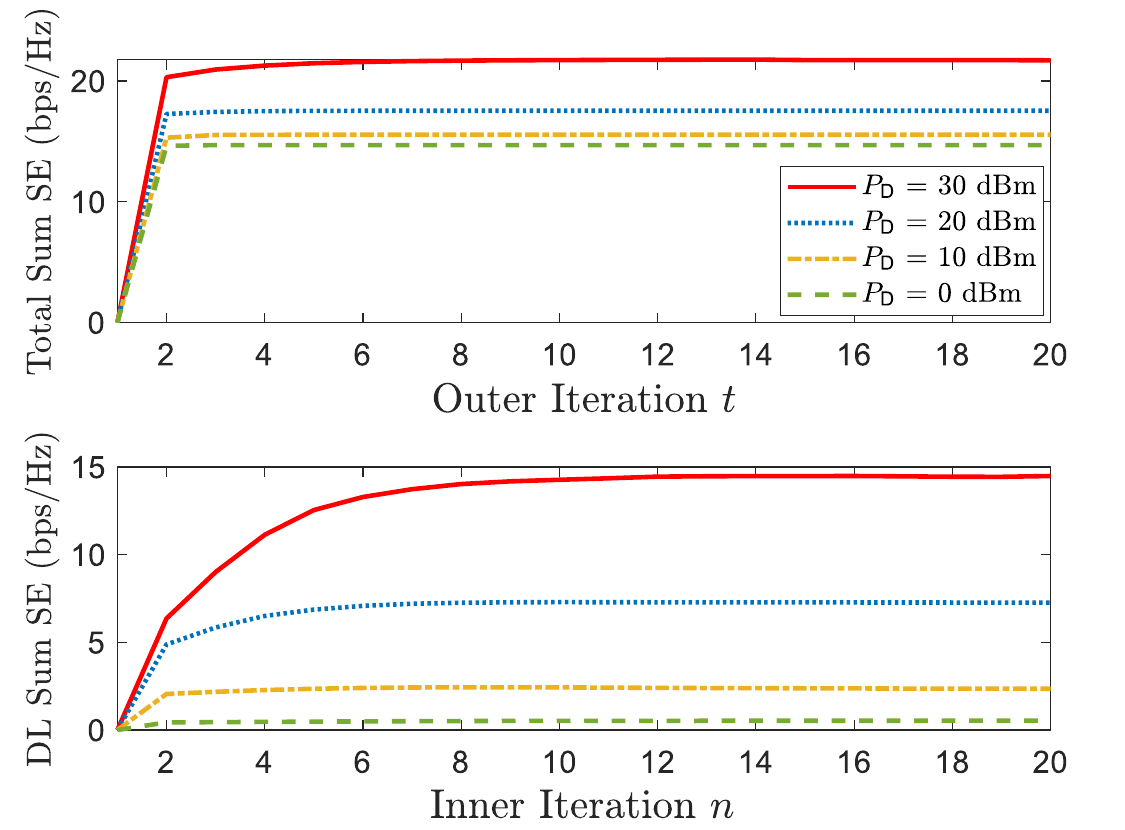}
    \caption{The convergence of the outer and inner iterations for $N_t=N_r=16$, $K_{\sf D}=K_{\sf U}=4$, $P_{\sf U}=23\,\mathrm{dBm}$, $\kappa_{\sf a}=-60\,\mathrm{dB}$, and  $b_{{\sf ADC},n}=b_{{\sf DAC},m}=7$ bits.}
    \label{fig:DL_convergence}
        \vspace{-1em}
\end{figure}
%%%%%%%%%%%%%%%%%%%%%%%%%%%%%%%%%%%%%%%
Fig.~\ref{fig:DL_convergence} shows the convergence of the proposed method by varying the AP's maximum transmit power.
In the simulations, the variables obtained in the last inner iteration serve as the initial value of the current outer iteration.
The results indicate that the proposed method converges for the considered systems.
In particular, as $P_{\sf D}$ increases, more inner iterations are required for convergence, while the number of outer iterations for convergence remains nearly constant.
This result demonstrates the fast convergence of the proposed method.

%%%%% Extension: Energy Efficiency %%%%%
\subsection{Energy Efficiency}
\label{subsec:EE}
In this subsection, we assess the EE performance of the proposed method.
According to \cite{choi:twc:2022}, the normalized EE is computed using the derived expressions in \eqref{eq:R_Dk} and \eqref{eq:R_Uk} as 
\begin{align}
    \mathrm{EE} = \frac{\sum_{k=1}^{K_{\sf D}} R_{{\sf D},k}(\bW) + \sum_{k=1}^{K_{\sf U}} R_{{\sf U},k}(\bW;\bF)}{P_{\sf AP}+ P_{\sf UE}}, \label{eq:EE_formulated_objective}
\end{align}
where the unit of EE is $\mathrm{bits/Joule/Hz}$, $P_{\sf AP}$ is a AP total power consumption, and $P_{\sf UE}$ is a sum of total power consumption of DL and UL users.
Based on \cite{zhang:2017:performance, zhang:tcom:18}, we define $P_{\sf AP}$ as 
\begin{align}
    P_{\sf AP} = P^{\sf tx}_{\sf AP} + P^{\sf rx}_{\sf AP},
    \label{eq:P_AP}
\end{align}
where $ P^{\sf tx}_{\sf AP} = P_{\sf LO} + N_t(2P_{\sf LP} + 2P_{\sf M} + P_{\sf H} + \kappa^{-1}_{\sf PA} P_{\sf TX}) + 2N_t(P_{\sf AGC} + P_{\sf DAC}(b_{\sf DAC}, f_s)) + P_{\sf BB}$ and $P^{\sf rx}_{\sf AP} = P_{\sf LO} + N_r(2P_{\sf LP} + 2P_{\sf M} + P_{\sf H} + P_{\sf LNA}) + 2N_r(P_{\sf AGC} + P_{\sf ADC}(b_{\sf ADC}, f_r)) + P_{\sf BB}$.
Here, $\kappa_{\sf PA}$ is the power amplifier efficiency, and $P^{\sf tx}_{\sf AP}$ and $P^{\sf rx}_{\sf AP}$ denote a transmit and receive power consumption at the AP, respectively.
$P_{\sf TX}$, $P_{\sf LO}$, $P_{\sf LP}$, $P_{\sf M}$, $P_{\sf H}$, $P_{\sf DAC}$, $P_{\sf ADC}$, $P_{\sf LNA}$, $P_{\sf AGC}$, and $P_{\sf BB}$ are the power consumption of target transmit power, local oscillator, low-pass filter, mixer, $90^{\circ}$ hybrid with buffer, DAC, ADC, low noise amplifier, automatic gain control, and baseband processor, respectively \cite{choi:twc:2022, zhang:2017:performance}.
We set $P_{\sf TX} = P_{\sf D}$.
The converter power consumptions $P_{\sf DAC}$ and $P_{\sf ADC}$ (W) are \cite{zhang:2017:performance}:
\begin{align}
    P_{\sf DAC} &=1.5 \times 10^{-5}\cdot2^{b_{\sf DAC}} + 9 \times 10^{-12} \cdot f_s \cdot b_{\sf DAC}, \\
    P_{\sf ADC}&=c_e \cdot f_r \cdot 2^{b_{\sf ADC}},
\end{align}
where $c_e$ is the power consumption per conversion step, and $f_{s}$ and $f_{r}$ are sampling rates of AP's transmitter and receiver, respectively.
We use $c_e = 494 \, \mathrm{fJ}$ \cite{lee:procieee:08}, and $f_{s} = f_{r} = 500$ MHz which is critically sampled. 
From \cite{zhang:tcom:18}, we represent $P_{\sf UE}$ as
\begin{align}
    P_{\sf UE} = K_{\sf D} P^{\sf DL}_{\sf UE} + K_{\sf U} P^{\sf UL}_{\sf UE},
    \label{eq:P_UE}
\end{align}
where the total power consumption of DL and UL users are 
\begin{align}
    P^{\sf DL}_{\sf UE} =&\, P_{\sf LO}\! + \!2P_{\sf LP} \!+\! 2P_{\sf M} \!+\! P_{\sf H} \!+\! P^{\sf UE}_{\sf LNA} \!+\! 2P_{\sf AGC}\! +\! P_{\sf BB}, \label{eq:P_UE_2}\\
    P^{\sf UL}_{\sf UE} =&\, P_{\sf LO} \!+\! 2P_{\sf LP}\! +\! 2P_{\sf M}\! + \!P_{\sf H}\! + \!P^{\sf UE}_{\sf PA} \!+\! 2P_{\sf AGC} \!+\! P_{\sf BB}.
    \label{eq:P_UE_3}
\end{align}
We assume that the parameters (e.g., $P_{\sf LO}$, $P_{\sf LP}$, $P_{\sf M}$, $P_{\sf H}$, $P_{\sf AGC}$, and $P_{\sf BB}$) used to define the total power consumption for the AP are the same as \eqref{eq:P_UE_2} and \eqref{eq:P_UE_3}.
The power consumptions of the low noise amplifier at the DL user and the power amplifier at the UL user are $P^{\sf UE}_{\sf LNA}$ and $P^{\sf UE}_{\sf PA} = \kappa_{\sf PA}^{-1}P_{\sf U}$, respectively.  
We set the circuit power consumption according to \cite{choi:twc:2022, zhang:2017:performance}.

%%%%%%%%%%%%% EEvsbits %%%%%%%%%%%%%%
\begin{figure}[t]
    \centering
    \includegraphics[width=0.815\linewidth]{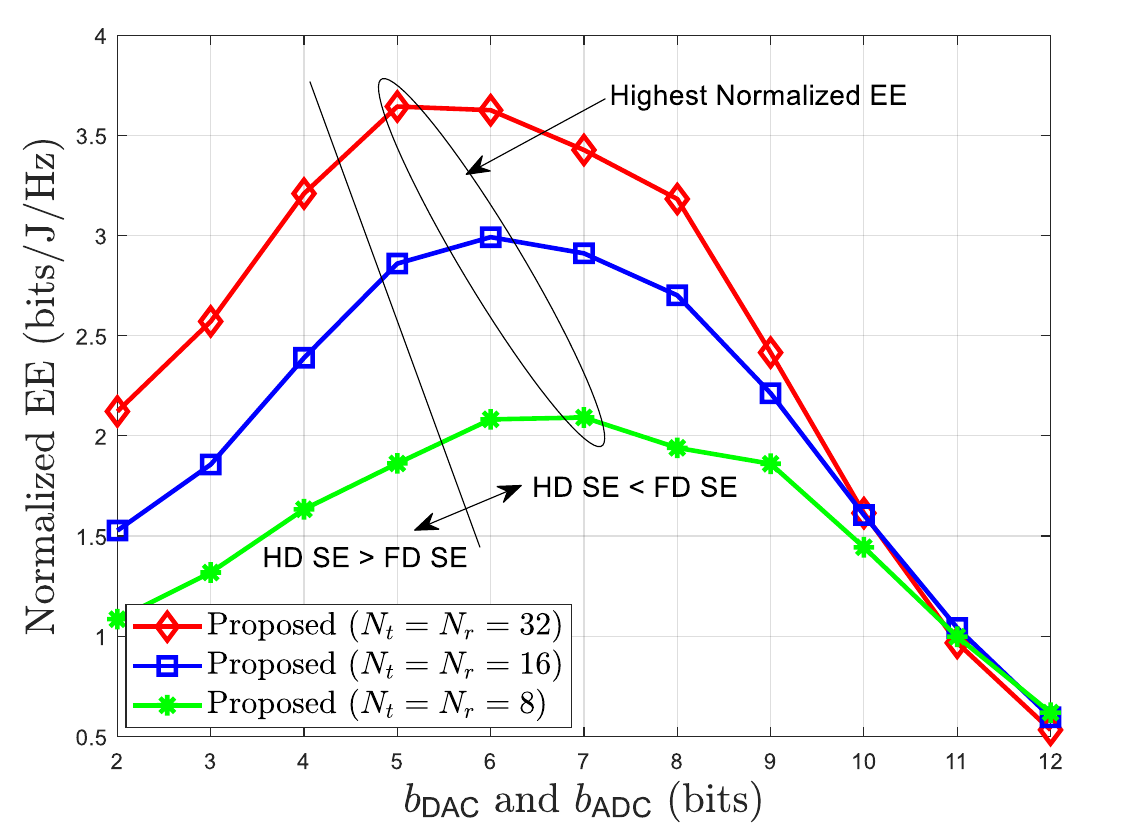}
    \caption{The normalized EE versus the ADC and DAC bits for $N_t=N_r \in \{8, 16, 32\}$, $K_{\sf D}=K_{\sf U}=4$, $P_{\sf D}=24\,\mathrm{dBm}$, $P_{\sf U}=23\,\mathrm{dBm}$, $\kappa_{\sf a}=-60\,\mathrm{dB}$.}
    \label{fig:EEvsbits}
    \vspace{-1em}
\end{figure}
%%%%%%%%%%%%%%%%%%%%%%%%%%%%%%%%%%%%%%
Fig.~\ref{fig:EEvsbits} shows the normalized EE versus the number of ADC and DAC bits.
A black line separates the region where the proposed algorithm yields the lower sum SE than the HD baseline (left) from the region where it outperforms the HD counterpart (right).
A key finding is that the normalized EE is maximized at the optimal resolution of about 6-7 bits.
Several prior studies on HD systems with low-resolution quantization demonstrate that using 4-5 bit quantizers is sufficient to achieve SE performance comparable to high-resolution quantization systems and the most energy-efficient strategy \cite{choi:twc:2022, studer2016quantized, verenzuela2016hardware, jacobsson2017throughput}.
However, as demonstrated in Fig.~\ref{fig:EEvsbits} as well as  Fig.~\ref{fig:SEvsADCbits}, more quantization bits are required to fully exploit the potential of the FD systems compared to the HD systems, which corresponds to the insights obtained from Section~\ref{sec:ADC}.

%%%%% Conclusion %%%%%
% Abstract에서 작성한 것과 같이, 1) 우리가 무엇을 분석했으며 어떤 finding이 있었고, 2) 이를 기반으로 어떤 알고리즘을 제안했으며, 3) 이론적인 분석을 numerical하게 분석하고 알고리즘의 성능 분석이 제대로 이루어졌다는 내용을 작성해야 함.
% 4) 마지막으로 우리 알고리즘의 우수성을 나타내는 문장(성능 분석에 어느 부분에 초점을 맞추어야 하는지)을 작성.
\section{Conclusion}
\label{sec:conclusion}
This paper presented a comprehensive analysis of quantized FD MU-MISO systems, with a particular focus on the implications of coarse quantization.
We first investigated the UL average SQNR under conventional linear precoding, demonstrating the impact of the residual analog SI and establishing the minimum ADC resolution required to satisfy specified SQNR targets.
The analyses reveal that the required number of ADC bits logarithmically increases with the residual analog SI, which hinders the FD system from employing low-resolution ADCs.
Motivated by these findings, we proposed an advanced SI-aware beamforming method to balance the SI suppression and DL SE maximization.
The validity of the analytical framework is corroborated by the strong consistency with the numerical evaluation of the proposed method, providing practical design guidelines for the optimization of FD systems in the presence of quantization errors.
Finally, it was  demonstrated numerically that 6-7 bit quantizers are desirable in maximizing both SE and EE, which is different from the HD systems where 4-5 bit quantizers are most efficient.

%%%%% APPENDIX %%%%%
\appendices
\section{Proof of Lemma \ref{lem:KKT}} \label{app:KKT}
We note that the power constraint in \eqref{eq:Rayleigh_constraint} can be ignored indeed since $\gamma_{{\sf D},k}(\bar {\bf v})$ and $\gamma_{{\sf U},k}(\bar {\bf v};{\bf F})$ are invariant up to the scaling of $\bar {\bf v}$. 
Then, the Lagrangian function of the problem is
\begin{align}
    \cL(\Bar{\bv}) = \mathrm{log}_2 \prod_{k=1}^{K_{\sf D}} \left(\frac{\Bar{\bv}^{\sf H} \bA_k \Bar{\bv}}{\Bar{\bv}^{\sf H} \bB_k \Bar{\bv}} \right)\prod_{k=1}^{K_{\sf U}} \left(\frac{\Bar{\bv}^{\sf H} \bC_{k} \Bar{\bv}}{\Bar{\bv}^{\sf H} \bD_{k} \Bar{\bv}} \right).\label{eq:Lagrangian}
\end{align}
We need to compute the derivative of \eqref{eq:Lagrangian} to obtain the first-order optimality condition.
Let us define $\lambda(\Bar{\bv})$ as
\begin{align}
    \lambda(\Bar{\bv}) =& \; \prod_{k=1}^{K_{\sf D}} \left( \frac{\Bar{\bv}^{\sf H}\bA_{k}\Bar{\bv}}{\Bar{\bv}^{\sf H} \bB_{k} \Bar{\bv}}\right)\prod_{k=1}^{K_{\sf U}} \left( \frac{\Bar{\bv}^{\sf H}\bC_{k}\Bar{\bv}}{\Bar{\bv}^{\sf H} \bD_{k} \Bar{\bv}}\right), 
\end{align}
and thus $\cL(\Bar{\bv})= \mathrm{log_2} \, \lambda(\Bar{\bv})$.

Now, we compute a derivative of $\cL(\Bar{\bv})$ which is denoted as
\begin{align}
    \frac{\partial \cL(\Bar{\bv})}{\partial \Bar{\bv}^{\sf H}} = \frac{1}{\lambda(\Bar{\bv})\mathrm{ln}\,2} \frac{\partial \lambda(\Bar{\bv})}{ \partial \Bar{\bv}^{\sf H}}. \label{eq:derivative_Lagrangian}
\end{align}
Thus, the derivative of $\lambda(\bar {\bv})$ is computed as
\begin{align}
    &\frac{\partial \lambda(\Bar{\bv})}{ \partial \Bar{\bv}^{\sf H}} = \label{eq:dlambda}
    \\
    &2\lambda(\Bar{\bv})\left(\sum_{k=1}^{K_{\sf D}} \left( \frac{\bA_{k}\Bar{\bv}}{\Bar{\bv}^{\sf H}\bA_{k}\Bar{\bv}} - \frac{\bB_{k}\Bar{\bv}}{\Bar{\bv}^{\sf H}\bB_{k}\Bar{\bv}} \right) + \sum_{k=1}^{K_{\sf U}} \left( \frac{\bC_{k}\Bar{\bv}}{\Bar{\bv}^{\sf H}\bC_{k}\Bar{\bv}} - \frac{\bD_{k}\Bar{\bv}}{\Bar{\bv}^{\sf H}\bD_{k}\Bar{\bv}} \right)\right). \nonumber
\end{align}
Setting \eqref{eq:dlambda} equal to zero with reorganization provides the stationary condition as
\begin{align}
    \label{eq:KKT_cond}
    &\lambda_{\sf num}(\Bar{\bv})\left(\sum_{k=1}^{K_{\sf D}} \frac{\bA_{k}}{\Bar{\bv}^{\sf H}\bA_{k}\Bar{\bv}} + \sum_{k=1}^{K_{\sf U}}  \frac{\bC_{k}}{\Bar{\bv}^{\sf H}\bC_{k}\Bar{\bv}} \right)\Bar{\bv} \nonumber  
    \\
    &= \lambda(\Bar{\bv})\lambda_{\sf den}(\Bar{\bv})\left(\sum_{k=1}^{K_{\sf D}}\frac{\bB_{k}}{\Bar{\bv}^{\sf H}\bB_{k}\Bar{\bv}}+\sum_{k=1}^{K_{\sf U}}\frac{\bD_{k}}{\Bar{\bv}^{\sf H}\bD_{k}\Bar{\bv}}\right)\Bar{\bv}.
\end{align}
Accordingly, we derive the necessary condition of the first-order optimality condition from \eqref{eq:KKT_cond} as
\begin{align}
    \bA_{\sf KKT}(\Bar{\bv})\Bar{\bv} = \lambda(\Bar{\bv})\bB_{\sf KKT}(\Bar{\bv})\Bar{\bv}, \label{eq:necessary_condition}
\end{align}
where $\bA_{\sf KKT}(\Bar{\bv})$ and $\bB_{\sf KKT}(\Bar{\bv})$ are defined in \eqref{eq:A_KKT} and \eqref{eq:B_KKT}, respectively. 
This completes the proof.
\qed

%%%%% Reference %%%%%
\bibliographystyle{IEEEtran}
\bibliography{bibtex}
\end{document}